\documentclass[]{aa}
\usepackage{graphicx}
\usepackage{times}
\usepackage{amsmath}
\usepackage{amssymb}

\begin{document}

\title{Dynamics of thick discs around Schwarzschild-de Sitter black
holes}

\author{Luciano	Rezzolla$^{(1, 2)}$, Olindo Zanotti$^{(1, 3)}$, 
	and Jos\'e A. Font$^{(3)}$
	} 

\institute{$^{(1)}$SISSA, International School for Advanced Studies,
	Via Beirut, 2-4 34014 Trieste, Italy \\
	$^{(2)}$INFN, Sezione di Trieste, Via A. Valerio, 2 34127
	Trieste, Italy \\
	$^{(3)}$Departamento de Astronom\'{\i}a y Astrof\'{\i}sica,
	Universidad de Valencia, %Edificio de Investigaci\'on. 
	Dr. Moliner 50, 
        46100 Burjassot, Spain 
	}

\titlerunning{Dynamics of thick discs around Schwarzschild-de Sitter
black holes}
\authorrunning{Zanotti, Rezzolla and Font} 

\date{Received Date~/~Accepted Date}

\abstract{We consider the effects of a cosmological constant on the
	dynamics of constant angular momentum discs orbiting
	Schwarzschild-de Sitter black holes. The motivation behind this
	study is to investigate whether the presence of a radial force
	contrasting the black hole's gravitational attraction can
	influence the occurrence of the runaway instability, a robust
	feature of the dynamics of constant angular momentum tori in
	Schwarzschild and Kerr spacetimes. In addition to the inner cusp
	near the black hole horizon through which matter can accrete onto
	the black hole, in fact, a positive cosmological constant
	introduces also an outer cusp through which matter can leave the
	torus without accreting onto the black hole. To assess the impact
	of this outflow on the development of the instability we have
	performed time-dependent and axisymmetric hydrodynamical
	simulations of equilibrium initial configurations in a sequence
	of background spacetimes of Schwarzschild-de Sitter black holes
	with increasing masses. The simulations have been performed with
	an unrealistic value for the cosmological constant which,
	however, yields sufficiently small discs to be resolved
	accurately on numerical grids and thus provides a first
	qualitative picture of the dynamics. The calculations, carried
	out for a wide range of initial conditions, show that the
	mass-loss from the outer cusp can have a considerable impact on
	the instability, with the latter being rapidly suppressed if the
	outflow is large enough.


} % end abstract

\maketitle

%=======================================================
\section{Introduction}
\label{I}
%=======================================================

	Relativistic accretion tori orbiting around stellar-mass black
holes have been the subject of renewed interest over the last few years
in connection with the different astrophysical scenarios where these
objects are expected to form, such as the core collapse of a massive star
leading to a ``failed" supernova explosion (a collapsar), or in the
catastrophic merger of two (unequal mass) neutron stars in a close binary
system. However, thick accretion discs are probably present at much
larger scales as well, surrounding quasars and other active galactic
nuclei, and feeding their central supermassive black holes. One of the
major issues about such systems concerns their dynamical stability. This
has important implications on the most favoured current models for the
central engines of $\gamma$-ray bursts, either collapsars or binary
neutron star mergers, for long and short bursts, respectively (see,
e.g. Meszaros 2002 for a recent review).

	Discs around black holes may suffer from a number of
instabilities produced either by axisymmetric or by non-axisymmetric
perturbations and further triggered by the presence of magnetic fields. A
type of instability that has been studied in a number of works and that
could take place when the discs are geometrically thick and axisymmetric
is the so-called {\it runaway instability} (see Font \& Daigne, 2002a;
Zanotti {\rm et al.} 2003 and references therein). To appreciate the
mechanism leading to the development of this instability, consider an
inviscid fluid torus with a vertical structure and internal pressure
gradients orbiting around a black hole (either Schwarzschild or Kerr). If
the fluid is non self-gravitating, it will be contained within
isopotential surfaces which generically possess a cusp on the equatorial
plane (Fishbone \& Moncrief, 1976; Kozlowski {\rm et al.} 1978;
Abramowicz {\rm et al.} 1978). As a result, material from the disc can
accrete onto the black hole through the cusp as the result of small
deviations from hydrostatic equilibrium.

	Any amount of matter lost by the disc and captured by the black
hole will increase its mass (and angular momentum), resulting in a
modification of the equipotential surfaces which may cause the cusp to
move deeper inside the torus more rapidly than the inner edge of the
torus. When this happens, additional disc material will be allowed to
fall into the black hole in an increasingly accelerated manner leading to
the runaway instability.

	Although this instability was first studied in the '80s
(Abramowicz {\rm et al.} 1983; Wilson, 1984), time-dependent
hydrodynamical simulations have been performed only recently, either with
SPH techniques and pseudo-Newtonian potentials (Masuda \& Eriguchi 1997;
Masuda, Nishida \& Eriguchi 1998), or with high-resolution
shock-capturing (HRSC hereafter) techniques in general relativity (Font
\& Daigne, 2002a Zanotti {\rm et al.} 2003). These investigations have
shown that, under the (idealised) assumption of constant specific angular
momentum distributions, relativistic tori around Schwarzschild and Kerr
black holes are generically unstable to the runaway instability, if non
self-gravitating.  The inclusion of more generic initial conditions,
however, can disfavour the occurrence of the instability. Recently, Font
\& Daigne (2002b) (see also Daigne \& Font, 2003) have shown through
numerical simulations that the runaway instability is suppressed when a
non-constant distribution of the angular momentum is assumed for the
torus (increasing as a power-law of the radius), a result which is in
agreement with studies based on a recent perturbative analysis (Rezzolla
{\rm et al.} 2003a; 2003b). While a similar stabilizing effect has been
shown to be provided by the black hole if this is rotating (Wilson, 1984;
Abramowicz {\rm et al.} 1998), Masuda \& Eriguchi (1997) were able to
show that the inclusion of the self-gravity of the torus effectively
favours the instability. Clearly, a final conclusion on the occurrence of
this instability has not been reached yet and will have to wait for fully
general relativistic simulations. However, the increasingly realistic
investigations performed recently have addressed several important
aspects and the prospects are that we may be close to reaching a detailed
description of the dynamics of the instability.

	A further physical process acting against the instability and
which has not been investigated so far, is provided by the existence of a
repulsive force pointing in the direction opposite to the black hole's
gravitational attraction. Such a force could disturb and even balance the
standard outflow of mass through the inner cusp, thus potentially
suppressing the runaway instability. As suggested recently by
Stuchl\'{\i}k {\rm et al.} (2000), such conditions could arise naturally
in a black hole spacetime with a positive cosmological constant, i.e. in
a Schwarzschild-de Sitter spacetime. In such a spacetime, in fact, a
second cusp appears in the outer parts of the equilibrium tori, near the
so-called ``static radius''. Assuming a value for the relict cosmological
constant $\Lambda \sim 10^{-56}$cm$^{-2}$ as deduced from recent
cosmological observations of the vacuum energy density (Krauss 1998) and
compatible with a sample of observational estimates provided by the
analysis of a large number of high redshift supernovae (Perlmutter {\it
et al.} 1999; Riess {\rm et al.} 1998), Stuchl\'{\i}k {\rm et al.}
(2000) find that the location of this outer cusp for the largest
stationary discs which can be built in a Schwarzschild-de Sitter
spacetime is at about $50-100$ kpc for supermassive black holes with
masses in the range $\sim 10^8 M_{\odot}-10^9 M_{\odot}$. As for the
inner one, a slight violation of the hydrostatic equilibrium at the outer
cusp would induce a mass outflow from the disc and away from the black
hole, which could affect the overall dynamics of the torus.

	However, this is not the only way in which a cosmological
constant could modify the dynamics of a disc orbiting around a
Schwarzschild-de Sitter black hole. As argued by Stuchl\'{\i}k {\it et
al.}  (2000), in fact, a cosmological constant could produce a sensible
modification in the accretion processes onto primordial black holes
during the very early stages of expansion of the Universe, when phase
transitions could take place, and the effective cosmological constant can
have values in many orders exceeding its present value (Kolb \& Turner
1990). Furthermore, a positive cosmological constant could also result
into strong collimation effects on jets escaping along the rotation axis
of the central black hole (Stuchl\'{\i}k {\rm et al.} 2000).

	The aim of this paper is to investigate one of these intriguing
possibilities through numerical simulations. More precisely, we present a
comprehensive study of the nonlinear hydrodynamics of constant angular
momentum relativistic tori evolving in a sequence of background
Schwarzschild-de Sitter spacetimes with increasing black holes
masses. Our study clarifies the dynamical impact of a mass outflow on the
occurrence of the runaway instability in such relativistic tori.

	We note that our setup will not be an astrophysically realistic
one. There are two important reasons for this. Firstly, given the present
estimates for the value of the cosmological constant and for the masses
of the supermassive black holes believed to exist in the centre of
galaxies and active galactic nuclei, we are still lacking sufficient
observational evidence that stationary thick accretion discs exist on
scales of about 100 kpc. Secondly, even assuming that such objects are
present within large galaxies, numerical calculations would have to face
the present computational limitations which make it extremely hard to
simulate accurately accretion discs with very low rest-mass densities and
over such length scales. As a result, we will adopt a value for the
cosmological constant that is unrealistically high. This yields discs
with radial extents that are sufficiently small to be evolved numerically
with satisfactory accuracy and provides a first qualitative description
of the role that a cosmological constant could play on the dynamics of
relativistic tori. In addition to this, the calculations reported here
also offer a way of assessing how the self-gravity of the torus, which is
basically contrasting the gravitational attraction of the black hole,
could modify the overall inertial balance. This will provide a useful
insight when fully relativistic calculations solving for the Einstein
equations coupled to a self-gravitating matter source will be performed.

	The organization of the paper is as follows. In Sect.~\ref{II}
we briefly review the main properties of relativistic tori in a
Schwarzschild-de Sitter spacetime. Next, in Sect.~\ref{III} we present
the hydrodynamics equations and the numerical methods implemented in our
axisymmetric evolution code. The material presented in
this Section is
rather limited, since the details have previously been reported in a
number of papers. The last part of this Section is devoted to a
discussion of the initial data we use for the simulations. The numerical
results are then described in Sect.~\ref{V} and, finally,
Sect.~\ref{VI} contains our conclusions.  Throughout the paper we use a
space-like signature $(-,+,+,+)$ and a system of geometrized units in
which $c = G = 1$. The unit of length is chosen to be the gravitational
radius of the black hole, $r_{\rm g} \equiv G M/c^2$, where $M$ is the
mass of the black hole. Greek indices run from 0 to 3 and Latin indices
from 1 to 3.

%=========================================================================
\section{Stationary configurations in a Schwarzschild-de Sitter spacetime}
\label{II}
%=========================================================================

	Building on a wide literature discussing equilibrium
configurations of perfect fluid relativistic tori orbiting around
Schwarzschild or Kerr black holes, Stuchl\'{\i}k {\it et~al.} (2000) have
recently extended these results to the case of a Schwarzschild-de Sitter
black hole. In spherical coordinates $(t,r,\theta,\phi)$ the line element
of this spacetime reads
\begin{eqnarray}
\label{metric}
&& ds^2=-\left(1-\frac{2M}{r}-y\frac{r^2}{M^2}\right)dt^2+
\nonumber \\ 
& & \hskip 0.25 truecm \left(1-\frac{2M}{r}
	- y \frac{r^2}{M^2}\right)^{-1}
	dr^2  + r^2(d\theta^2+\sin^2\theta d\phi^2) \ ,
\end{eqnarray}
where $M$ is the mass of the black hole and the cosmological constant
$\Lambda$ is incorporated in the dimensionless parameter $y$ defined as
\begin{equation}
y \equiv\frac{1}{3}\Lambda M^2 \ .
\end{equation}
This parameter has to be smaller than a critical value $y < y_c \equiv
1/27$ in order to produce static regions of the spacetime where
equilibrium configurations can be found. We note that a negative
cosmological constant, corresponding to a Schwarzschild-anti de Sitter
black hole, does not introduce new qualitative features in the
development of the runaway instability when compared to a Schwarzschild
spacetime and will not be considered here.  In what follows we briefly
recall the main features of stationary configurations in a
Schwarzschild-de Sitter spacetime, referring the reader to the work of
Stuchl\'{\i}k {\rm et al.} (2000) for an exhaustive discussion.

	The perfect fluid with four-velocity $u^{\mu}$ is described by
the usual stress-energy tensor
\begin{equation}
\label{stress-tensor}
T^{\mu\nu}\equiv (e+p)u^\mu u^\nu+p g^{\mu\nu}
	= \rho h u^\mu u^\nu+p g^{\mu\nu} \ ,
\end{equation}
where $g^{\mu\nu}$ are the coefficients of the metric (\ref{metric})
in which, however, the black hole mass $M$ could be a function of time to account
for the mass accreted onto the black hole [cf. the
discussion of Eq.~(\ref{met_up}) below]. 
The fluid variables $e$, $p$, $\rho$, and $h =
(e+p)/\rho$ are the proper energy density, the isotropic
pressure, the rest mass density, and the specific
enthalpy, respectively.  An equation of state (EOS) of
polytropic type, $p = \kappa \rho^\gamma =
\rho\epsilon(\gamma-1)$, completes the thermodynamical
description of the fluid. Here, $\kappa$ is the
polytropic constant, $\gamma$ is the adiabatic index and
$\epsilon=e/\rho - 1$ is the specific internal energy. As
shown by Kozlowski {\rm et al.}  (1978) (see also
Fishbone \& Moncrief, 1976), the pressure gradients can
balance the gravitational and centrifugal forces,
allowing for the existence of stationary configurations
of matter in non-geodesic circular motion and contained
within closed ``constant pressure'' equipotential
surfaces. Under the conditions of hydrostatic equilibrium
and of axisymmetry (i.e. $\partial_t = \partial_{\phi} =
0$) the relativistic Euler equations for a fluid with
four-velocity $u^{\alpha} = (u^t, 0, 0, u^{\phi})$ take
the simple Bernouilli-type form 
\begin{equation}
\label{bernoulli}
\frac{\nabla_i p}{e+p} = - \nabla_i W +
	\frac{\Omega \nabla_i \ell}{1- \Omega \ell}\ ,
	\hspace{2cm} i=r,\theta,
\end{equation} 
where $W = W(r,\theta) \equiv \ln(u_t)$ is the effective potential,
$\ell\equiv -u_\phi/u_t$ is the specific angular momentum, and
$\Omega\equiv u^\phi/u^t$ is the coordinate angular velocity as measured
by an observer near the static radius, where the spacetime geometry is
very close to a flat one (Stuchl\'{\i}k {\rm et al.} 2000). Note that an
explicit relation exists between the angular velocity and the specific
angular momentum, which is given by $\Omega=-\ell (g_{tt}/g_{\phi\phi})$.

	Once $M$ and $\Lambda$ have been prescribed, the explicit
expression for the potential $W(r,\theta)$ in the Schwarzschild-de Sitter 
spacetime is simply given by (Stuchl\'{\i}k {\rm et al.} 2000)
\begin{equation}
\label{W_r_t}
W (r, \theta) = \frac{1}{2}\ln \left[ \frac{ (1 - 2
        {M}/{r} - y {r^2}/{M^2}) r^2
        \sin^2\theta} {r^2 \sin^2\theta - (1 - 2 {M}/{r} -
        y {r^2}/{M^2}) \ell^2}      \right]\ .
\end{equation} 

\begin{figure*}
\begin{center}
\includegraphics[width=8.7cm,angle=0]{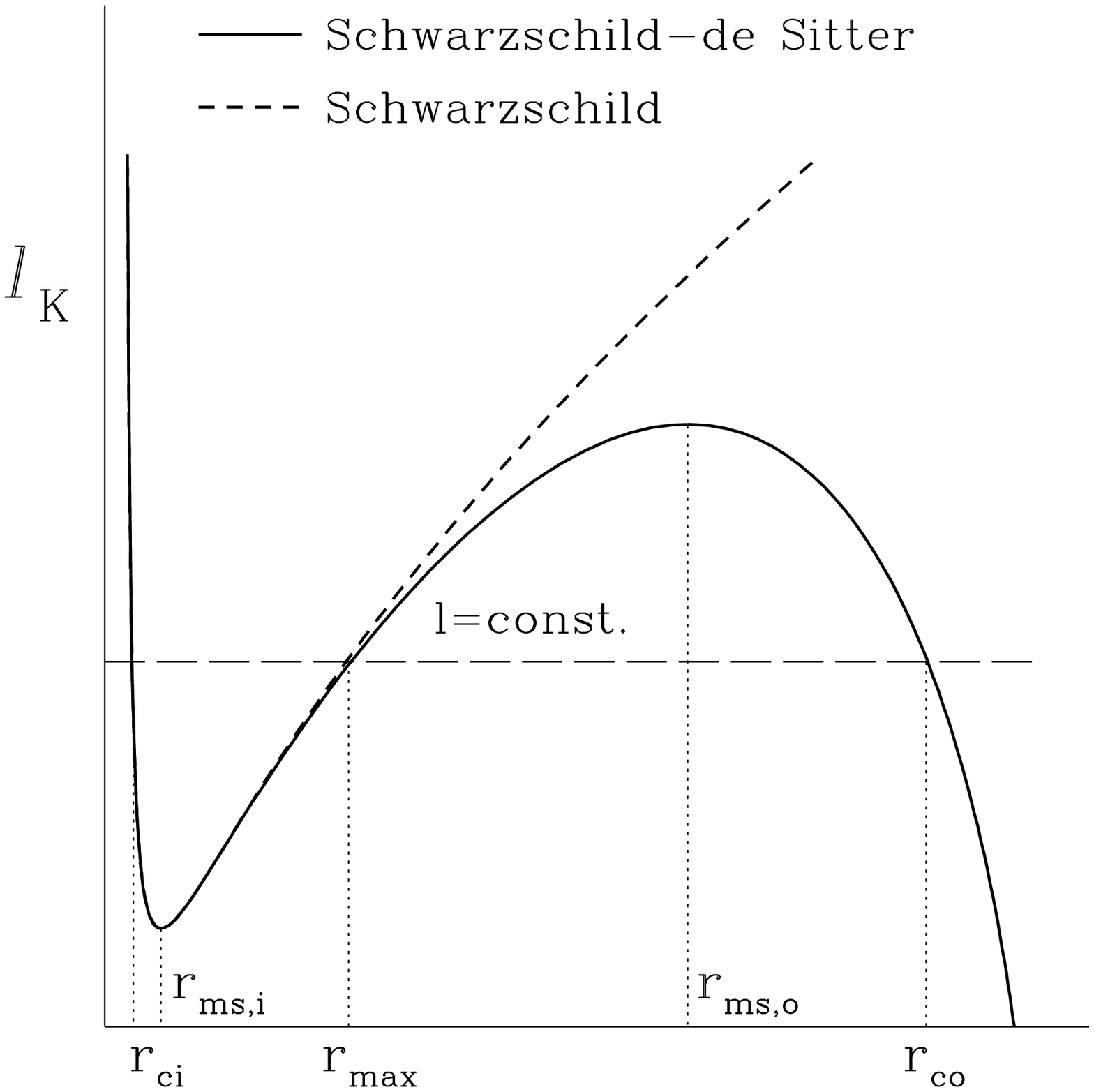}
\hspace{0.05truecm}
\includegraphics[width=8.7cm,angle=0]{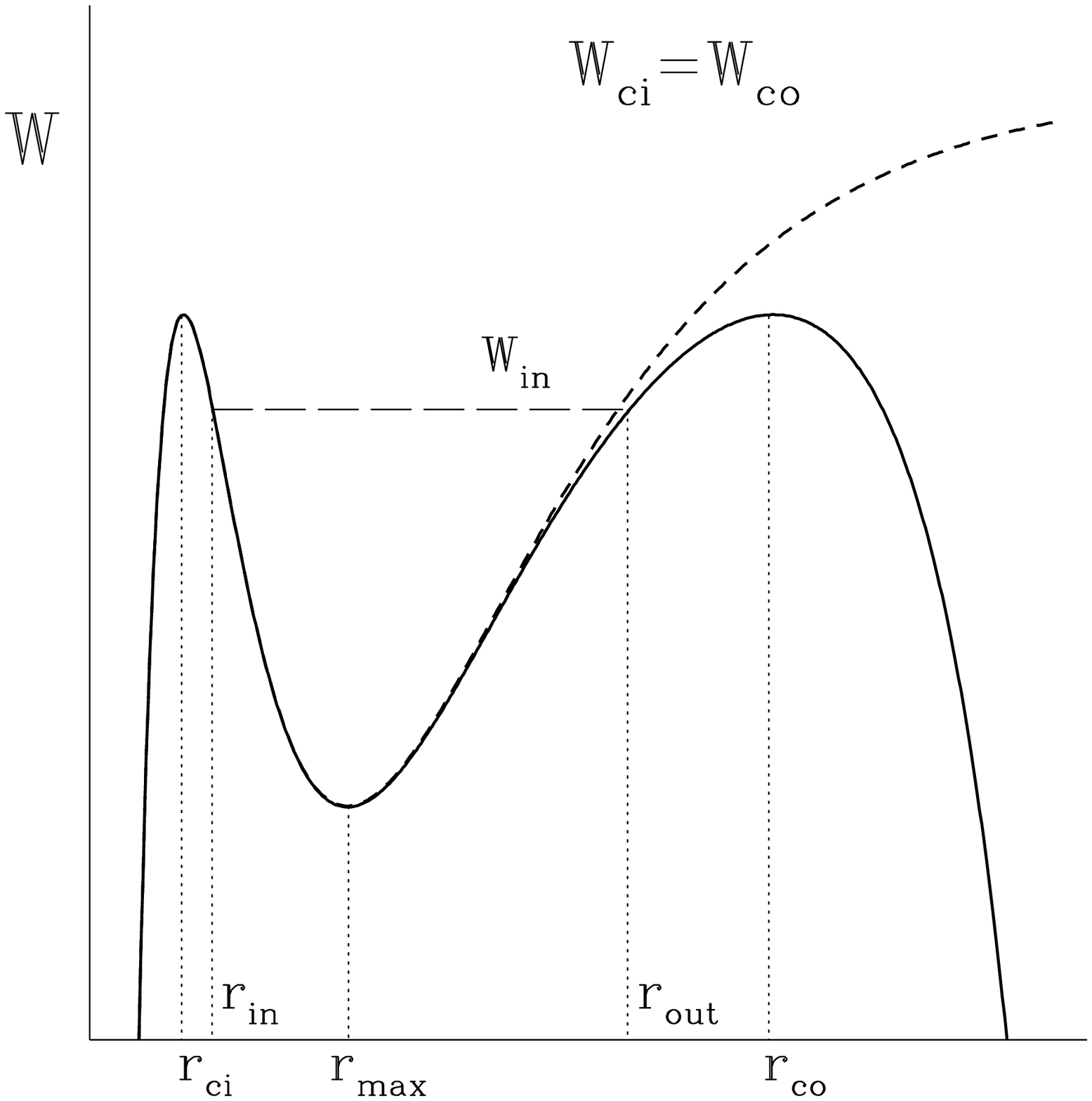}
\caption{Schematic diagram for the Keplerian specific angular momentum
$\ell_{_K}$ in a Schwarzschild-de Sitter spacetime (left panel) and the
corresponding effective potential $W$ (right panel) once a constant value
for $\ell$ has been chosen (cf. Fig.\ref{fig1}). The figure reports the
different radial locations that are relevant for our discussion: the
inner and outer cusp points $r_{\rm ci}, r_{\rm co}$, the inner and outer
radii for the torus $r_{\rm in}, r_{\rm out}$, the inner and outer
marginally stable orbits $r_{\rm ms,i}, r_{\rm ms,o}$, and the location
of the maximum pressure in the torus $r_{\rm max}$ (see text for
details). Note that $r_{\rm ci}, r_{\rm co}$ and $r_{\rm max}$ are
determined once a value for the constant specific angular momentum has
been chosen (this is shown with the long-dashed line in the left panel)
and that the inner and outer radii need not coincide with the
corresponding locations of the cusps but are set by the value chosen for
the potential $W_{\rm in}$ (this is shown with the long-dashed line in
the right panel). Reported for comparison with a short-dashed line are
$\ell_{_K}$ and $W$ in a Schwarzschild spacetime; the radial coordinate
is shown on a logarithmic scale.}
\label{fig0}
\end{center}
\end{figure*}

	It is apparent from Eq.~(\ref{bernoulli}) that the simplest and
indeed best studied configurations are obtained when the distribution of
specific angular momentum $\ell$ is prescribed to be constant. In this
case, which is also the one considered here, $W(r,\theta)$ shows three
local extrema in the equatorial plane, one more than in the case of a
Schwarzschild spacetime (cf. Fig.~\ref{fig0}). Ordering these points with
increasing values of $r$, the first extremum corresponds to the position
of the inner cusp, $r_{\rm{ci}}$, where the equipotential surface has a
self-crossing point in the $(r,\theta)$ plane. The second extremum
corresponds to the position of the ``centre'' of the torus, $r_{\rm
max}$, where the internal pressure of the torus has its maximum. The
third extremum, finally, marks the position of the outer cusp,
$r_{\rm{co}}$, which is not present in the Schwarzschild spacetime and
represents the distinctive contribution of the non-zero cosmological
constant (see Fig.~\ref{fig0} for a schematic diagram). Note that for a
configuration with constant $\ell$, a local extremum of $W$ is also a
point of vanishing pressure gradients [cf. eq. (\ref{bernoulli})].  As a
result, the fluid motion on a circular orbit of radius $r_{\rm{ci}}$,
$r_{\rm max}$ or $r_{\rm{co}}$ is a purely geodetic one, with the
specific angular momentum being given by the Keplerian one $\ell_{_K}$
for a point-like particle at that radius. In a Schwarzschild-de Sitter
spacetime this is given by
\begin{equation}
\label{l_kep}
\ell^2_{_K}(r,y) \equiv \frac{r^3M(1-y
	{r^3}/{M^3})}{(r - 2M - y {r^3}/{M^2})^2}\ .
\end{equation}

	If the black hole is a Schwarzschild one, then $y=0$ and the
admittable values for a constant distribution of specific angular
momentum that will lead to a torus of finite size are
\begin{equation}
\label{ell_intrvl}
\ell_{\rm ms} < \ell < \ell_{\rm mb} \ ,
\end{equation}
where $\ell_{\rm ms} = 3\sqrt{6}/2$ and $\ell_{\rm mb} = 4$ represent the
Keplerian specific angular momentum at the marginally stable and at the
marginally bound orbits, respectively. In the case of a Schwarzschild-de
Sitter black hole, however, the allowed parameter space is more complex
since in this case one needs to take into account also the admittable
values for $y$. While a detailed discussion of this issue is provided by
Stuchl\'{\i}k {\rm et al.} (2000), we here remind that for $y\ne 0$, a
toroidal configuration of finite size is obtained for values of the
specific angular momentum that satisfy
\begin{equation}
\label{ell_intrvl_yneq0}
\ell_{\rm ms,i} < \ell < \ell_{\rm ms,o} \ ,
\end{equation}
where now $\ell_{\rm ms,i}$ and $\ell_{\rm ms,o}$ represent the values of
the specific angular momentum at the inner and outer marginally stable
radii, respectively, and correspond to the local minimum and maximum of
the Keplerian specific angular momentum given by Eq.~(\ref{l_kep})
(cf. Fig.~\ref{fig0}). Hence, $\ell_{\rm ms,i}$ and $\ell_{\rm ms,o}$
provide the minimum and maximum values of $\ell(r,y)$ for which stable
Keplerian circular orbits exist.  Note that the very existence of these
stable circular orbits depends on the values of $y$ and, in particular,
these can be found only if $y < y_{\rm ms} =12/15^4 \sim 2.37\times
10^{-4} < y_c$.

	In other words, given a value for the dimensionless parameter
${\bar y} < y_{\rm ms}$, circular orbits can exist for $\ell_{\rm
ms,i}({\bar y}) < \ell < \ell_{\rm ms,o}({\bar y})$. Clearly, these
orbits will be at radii smaller than the static radius $r_s\equiv {\bar
y}^{-1/3}$, at which the angular momentum (\ref{l_kep}) is zero and where
the gravitational attraction of the black hole is exactly balanced by the
cosmological repulsion\footnote{The Keplerian specific angular momentum
is not well defined for $r>r_s$, where its square is negative.}.

	Finally, it is worth reminding that for point-like particles, the
decreasing part of $\ell^2_{_{K}}$ (i.e. for $r<r_{\rm{ms,i}}; \; r
>r_{\rm{ms,o}}$) corresponds to unstable circular geodesics, while the
increasing part (i.e. for $r_{\rm{ms,i}} < r <r_{\rm{ms,o}}$) corresponds
to stable circular geodesics. For a perfect fluid, however, internal
pressure gradients allow the torus to occupy a region which would be
geodesically unstable for point-like particles. A schematic diagram
showing the behaviour of the Keplerian specific angular momentum
$\ell_{_K}$ and the corresponding effective potential $W$ in a
Schwarzschild-de Sitter spacetime, as well as the locations of the
different radii that have been discussed so far is presented in
Fig.~\ref{fig0}. The Schwarzshild case is also displayed for comparison
with a dashed line.

	Once a value for the cosmological parameter and for the angular
momentum have been fixed, Eq.~(\ref{bernoulli}), supplemented by the
polytropic EOS, can be integrated analytically for any $r \leq r_s$, to
yield the rest-mass density distribution inside the torus
\begin{equation}
\label{density}
\rho(r,\theta) = \left\{\frac{\gamma-1}{\kappa \gamma}
	\left[\exp({W_\mathrm{in}-W})-1\right]\right\}^{1/(\gamma-1)} \ ,
\end{equation}
where $W_{\rm in} \equiv W(r_{\rm in},\pi/2)$ and $r_{\rm{in}}$ is the
inner edge of the torus. The latter is assumed to be a free parameter and
is effectively controlled by the potential gap $\Delta W_{\rm i} \equiv
W(r_{\rm in},\pi/2) - W(r_{\rm ci},\pi/2)$.

	Hereafter, we will focus on tori built in a parameter space that
is smaller than the one discussed so far. In particular, we will consider
tori with constant specific angular momentum in the range
$\ell_{\rm{ms,i}} \leq \ell\leq\ell_{\rm{ph}} < \ell_{\rm ms,o}$, where
$\ell^2_{\rm{ph}} \equiv r^3/(r - 2M - y r^3/M^2)$ is the angular
momentum of the unstable photon circular geodesic (Stuchl\'{\i}k {\it et
al.} 2000). Furthermore, the hydrodynamical evolution of these tori will
be followed in Schwarzschild-de Sitter spacetimes with dimensionless
cosmological constant $0 \leq y \leq y_e = 1/118125 \sim 8.46\times
10^{-6}$, where $y_e$ corresponds to the value of $y$ for which the
minimum of $\ell_{\rm{ph}}$ is equal to the Keplerian angular momentum of
the outer marginally stable orbit (Stuchl\'{\i}k {\rm et al.} 2000).

%=======================================================
\section{Hydrodynamic Equations and Numerical Approach}
\label{III}
%=======================================================

%%%%%%%%%%%%%%%%%%%%%%%%%%%%%%%
\subsection{Basic Equations}
%%%%%%%%%%%%%%%%%%%%%%%%%%%%%%%

	Suitable initial data for the torus (see Sect.~\ref{IV} for
specific details) are evolved forward in time using a two-dimensional
(2D) general relativistic hydrodynamics code. The equations for a perfect
fluid are implemented using the formulation developed by Banyuls {\it et
al.}  (1997) in which, adopting a $3+1$ decomposed spacetime, the
covariant conservation laws of the density current and of the
stress-energy tensor are recast into a first-order, flux-conservative
system of the type
\begin{eqnarray}
\label{fcf}
\frac{\partial {\bf U}}{\partial t} +
	\frac{\partial [\alpha {\bf F}^{r}]}{\partial r} +
	\frac{\partial [\alpha {\bf F}^{\theta}]}{\partial \theta}
	 = {\boldsymbol {\cal S}} \ ,
\label{system}
\end{eqnarray}
where $\alpha\equiv\sqrt{-g_{00}}$ is the lapse function of the
Schwarzschild-de Sitter metric and ${\bf U} \equiv (D, S_r, S_{\theta},
S_{\phi})$ is the state-vector of {\it ``conserved''} variables. These
represent the relativistic rest-mass density and the relativistic momenta
in the three coordinate directions, respectively. The conserved variables
are expressed in terms of the {\it ``primitive''} variables $\rho,
v_{i},$ and $\epsilon$ through the relations $D \equiv \rho \Gamma$ and
$S_j \equiv \rho h \Gamma^2 v_j$, where the covariant components of the
three-velocity $v_i$ can be obtained from the contravariant components
$v^i\equiv u^i/\alpha u^t$ through the coefficients of the spatial
3-metric $\gamma_{ij}$, i.e. $v_i=\gamma_{ij}v^j$. The Lorentz factor
$\Gamma$ is assumed to be measured by a local static observer and is
given by $\Gamma\equiv\alpha u^t = (1-v^2)^{-1/2}$, with $v^2 \equiv
\gamma_{ij}v^i v^j$. Finally, the vectors ${\bf F}^{i}$ and ${\boldsymbol
{\cal S}}$ appearing in Eq.~(\ref{fcf}) represent the fluxes and sources
of the evolved quantities, respectively, the latter being due entirely to
the spacetime curvature. The explicit expressions of these terms can be
easily generalized from the Schwarzschild case and are
\begin{eqnarray}
{\bf F}^{r}({\bf w})  =  \left(D v^{r},
 S_r v^{r} + p, S_{\theta} v^{r}, S_{\phi} v^{r} \right) \ ,
\end{eqnarray}
\noindent
\begin{eqnarray}
{\bf F}^{\theta}({\bf w})  =   \left(D v^{\theta}\ , 
S_{r} v^{\theta}\ , 
S_{\theta} v^{\theta} + p\ ,
S_{\phi} v^{\theta}\right)\ ,
\end{eqnarray}
\noindent
\begin{eqnarray}
{\boldsymbol {\cal S}}({\bf w}) = 
	({\cal S}_1,{\cal S}_2,{\cal S}_3,{\cal S}_4)\ ,
\end{eqnarray}
where
\begin{eqnarray}
{\cal S}_1 & \equiv & D({\cal A} v^r - {\cal B} v^\theta) \ ,
\\ \nonumber \\
{\cal S}_2 & \equiv & -\left(\frac{M}{\alpha r^2}-
	\frac{r y}{\alpha M^2}\right)(\tau + D) -
	\alpha\cot\theta S_r v^\theta + \nonumber \\ 
&& \hskip 3.0 cm \frac{\alpha}{r}(S_\theta
	v^\theta + S_\phi v^\phi - 2 S_r v^r) \ ,
\\ \nonumber \\
{\cal S}_3 & \equiv & S_\theta ({\cal A} v^r - {\cal B}
v^\theta) + \alpha r^2 S^\phi v^\phi\sin\theta \cos\theta \ , 
\\ \nonumber \\
{\cal S}_4 & \equiv & S_\phi({\cal A}v^r - {\cal B}v^\theta) \ ,
\end{eqnarray}
\noindent
and where
\begin{eqnarray}
{\cal A} & \equiv & \frac{M}{\alpha r^2} - \frac{2\alpha}{r} -
\frac{r y}{\alpha M^2}\ , 
\\ \nonumber \\
{\cal B} & \equiv & \alpha \cot\theta \ .
\end{eqnarray}
It should be noted that since the dynamical evolution is assumed to be
adiabatic, no evolution equation for the relativistic total energy
density $\tau\equiv \rho h \Gamma^2 - p -D$ is solved here.

%%%%%%%%%%%%%%%%%%%%%%%%%%%%%%%
\subsection{Numerical approach}
%%%%%%%%%%%%%%%%%%%%%%%%%%%%%%%

	The numerical code we use is the same employed by Zanotti {\it et
al.} (2003) to study the dynamics of constant angular momentum
relativistic discs around a Schwarzschild black hole. For the present
investigation the code has been extended to account for the modifications
introduced by the Schwarzschild-de Sitter geometry. The general
relativistic hydrodynamics equations are solved by means of a HRSC scheme
based on Marquina's flux formula (see e.g. Font, 2000 for a review of
these methods in numerical general relativistic hydrodynamics).

	In order to cover optimally the large spatial extent of the
equilibrium configurations and yet reach a satisfactory spatial
resolution in the regions closer to the two cusps where the fluid motion
needs to be calculated most accurately, we have introduced an important
technical modification in the handling of the radial-coordinate
grid. More precisely, we use a non-uniform radial grid with a logarithmic
spacing, which is double-varied in the vicinities of the inner and outer
cusps.  The coordinate mapping used for this purpose is reminiscent of a
tortoise coordinate mapping but it has been extended to a
Schwarzschild-de Sitter metric as
\begin{equation}
r_{\ast}= \pm \int \left(1-\frac{2M}{r}-y \frac{r^2}{M^2}\right)^{-1}dr
	\ ,
\end{equation}
where the $\pm$ sign distinguishes whether the mapping is made for
increasing or decreasing values of the coordinate $r$, respectively.

	 As a result of this mapping, a radial grid of $N_r=300$ zones
allows to cover a spatial domain going from $r_{_{\rm{MIN}}}= 2.1$ to
$r_{_{\rm{MAX}}} = 100$ with a minimum radial spacing of the innermost
part of the grid $\Delta r =10^{-4}$ and, correspondingly, a minimum
radial spacing $\Delta r=10^{-3}$ for the outermost part of the radial
grid. The two grids join smoothly at $r=48.6$, where the resolution is
$\Delta r=2.72$. The angular grid, on the other hand, is more
straightforward to build and consists of $N_{\theta}=70$ equally spaced
zones extending from $0$ to $\pi$ (cf. Zanotti {\rm et al.} 2003)

	As in Zanotti {\rm et al.} (2003), a low density ``atmosphere" is
introduced in those parts of the numerical domain outside the torus. The
initial atmosphere model chosen corresponds to the spherically symmetric
accretion solution of non-interacting test fluid particles. The maximum
density of the atmosphere is typically $5$ to $6$ orders of magnitude
smaller than the density at the centre of the torus. In all of the
validating tests performed, the hydrodynamical evolution of the torus was
found to be unaffected by the presence of this atmosphere, which is
evolved as the bulk of the fluid.

	Finally, the mass outflows at the innermost and outermost radial
points are computed respectively as
\begin{equation}
\label{mdot_i}
\dot{m}_{\rm i}(r_{_{\rm MIN}})\equiv - 2\pi\int_0^{\pi} 
	\sqrt{-g} D v^r d\theta 
	\Big\vert_{r_{_{\rm MIN}}}\ ,
\end{equation}
and 
\begin{equation}
\label{mdot_o}
\dot{m}_{\rm o}(r_{_{\rm MAX}})\equiv 2\pi\int_0^{\pi} \sqrt{-g} D v^r d\theta 
	\Big\vert_{r_{_{\rm MAX}}}\ ,
\end{equation}
where $g$ is the determinant of the metric and $\sqrt{-g}=r^2\sin\theta$.

\begin{table*}
\begin{center}
\caption{Main properties of the tori considered in the numerical
calculations.  From left to right the columns report: the name of the
model, the specific angular momentum $\ell$ (normalized to $M$), the
polytropic constant $\kappa$, the inner and outer cusps of the torus,
$r_\mathrm{ci}$ and $r_\mathrm{co}$, the radial position of the pressure
maximum $r_\mathrm{max}$ (all radii are in units of the gravitational
radius $r_\mathrm{g}$), the potential gaps $\Delta W_{\rm{i}}\equiv
W_{\rm{in}}-W_{\rm{ci}}$ and $\Delta W_{\rm{o}}\equiv
W_{\rm{in}}-W_{\rm{co}}$, where $W_{\rm{in}}$ is the potential at the
inner edge of the disc. The last column reports the orbital period at the
centre of the torus, $t_\mathrm{orb}$, expressed in milliseconds.  All of
the models share the same value of the cosmological parameter
$y=10^{-6}$, the same mass for the black hole, $M=10M_{\odot}$, the same
adiabatic index $\gamma=4/3$, and the same torus-to-hole mass ratio
$M_{\rm t}/M=0.2$.  }
\label{tab1}
\begin{tabular}{l|lc|ccc|cc|c}
\hline
& & & & & & & & \\
Model   & $\ell$    & $\kappa$ (cgs)  & $r_{\rm ci}$     &
$r_{\rm co}$ & $r_{\rm max}$  
& $\Delta W_{\rm i}$	 & $\Delta W_{\rm o}$ &  $t_{\rm orb}$ (ms)
\\
& & & & & & & & \\
\hline
$A_1$   & 3.84     & 8.970${\times} 10^{14}$ &4.419 &
	94.866 & 8.822  & 0.010 & -0.010 & 8.11 \\
$A_2$   & 3.84    & 2.568${\times} 10^{15}$  &4.419 &
	94.866 & 8.822  & 0.025 & 0.005  & 8.11  \\
$A_3$   & 3.84    & 4.372${\times} 10^{15}$  &4.419 &
        94.866 & 8.822  & 0.032 & 0.012  & 8.11 \\
\hline
$B_1$   & 3.94  &   2.295${\times} 10^{15}$ & 4.133&
	94.564 & 9.876 & 0.004 & 0.004 & 9.61 \\
$B_2$   & 3.94  &   3.775${\times} 10^{15}$ & 4.133&
        94.564 & 9.876 & 0.010  & 0.010  & 9.61 \\
$B_2$   & 3.94  &   6.740${\times} 10^{15}$ & 4.133&
	94.564 & 9.876 & 0.020  & 0.020  & 9.61 \\
\hline
$C_1$   & 4.00   &   3.025${\times} 10^{15}$ &4.000 & 
	94.373 & 10.489 & -0.007 & 0.007 & 10.51 \\
$C_2$   & 4.00   &   7.120${\times} 10^{15}$ &4.000 &
	94.373 & 10.489 & 0.007  & 0.021 & 10.51 \\
$C_2$   & 4.00   &   1.125${\times} 10^{16}$ &4.000 &
        94.373 & 10.489 & 0.020  & 0.034 & 10.51 \\
\hline
\end{tabular}
\end{center}
\end{table*}

	Note that the mass outflow given by Eq.~(\ref{mdot_i})
corresponds effectively to the mass accretion rate onto the black hole
and is used to account for the instantaneous increase of the black hole
mass at every time step. This, in turn, provides information about the
changes in the background spacetime, fundamental for the appearance of
the runaway instability (Font \& Daigne, 2002a; Zanotti {\rm et al.}
2003). As mentioned in the Introduction, since we neglect the
self-gravity of the torus, the hydrodynamics equations are solved in a
sequence of background Schwarzschild-de Sitter spacetimes with increasing
black hole masses. In practice, the spacetime evolution is achieved
through a remapping of the metric functions at each time level of the
type
\begin{equation}
\label{met_up}
g_{\mu \nu}(r,M^{n},y) \longrightarrow {\tilde
        g}_{\mu \nu}(r,M^{n+1},y) \ ,
\end{equation}
where $M^{n+1}=M^n + \Delta t \ \dot m_{\rm i}^n(r_{_{\rm MIN}})$ is the
mass of the black hole at the new timelevel $t^{n+1}$. A detailed
discussion on the validity of this approximation can be found in Font \&
Daigne (2002a) and in Zanotti {\rm et al.} (2003). The prescription
(\ref{met_up}) is justified and can be regarded as a very good
approximation when the variation of the black hole mass per unit time,
$\dot m_{\rm i}^n$, is very small. This is certainly the case for the
small disc-to-hole mass ratios $M_{\rm t}/M$ considered here
(cf. Table~\ref{tab1}).

\begin{figure*}
\begin{center}
\includegraphics[width=6.75cm,angle=0]{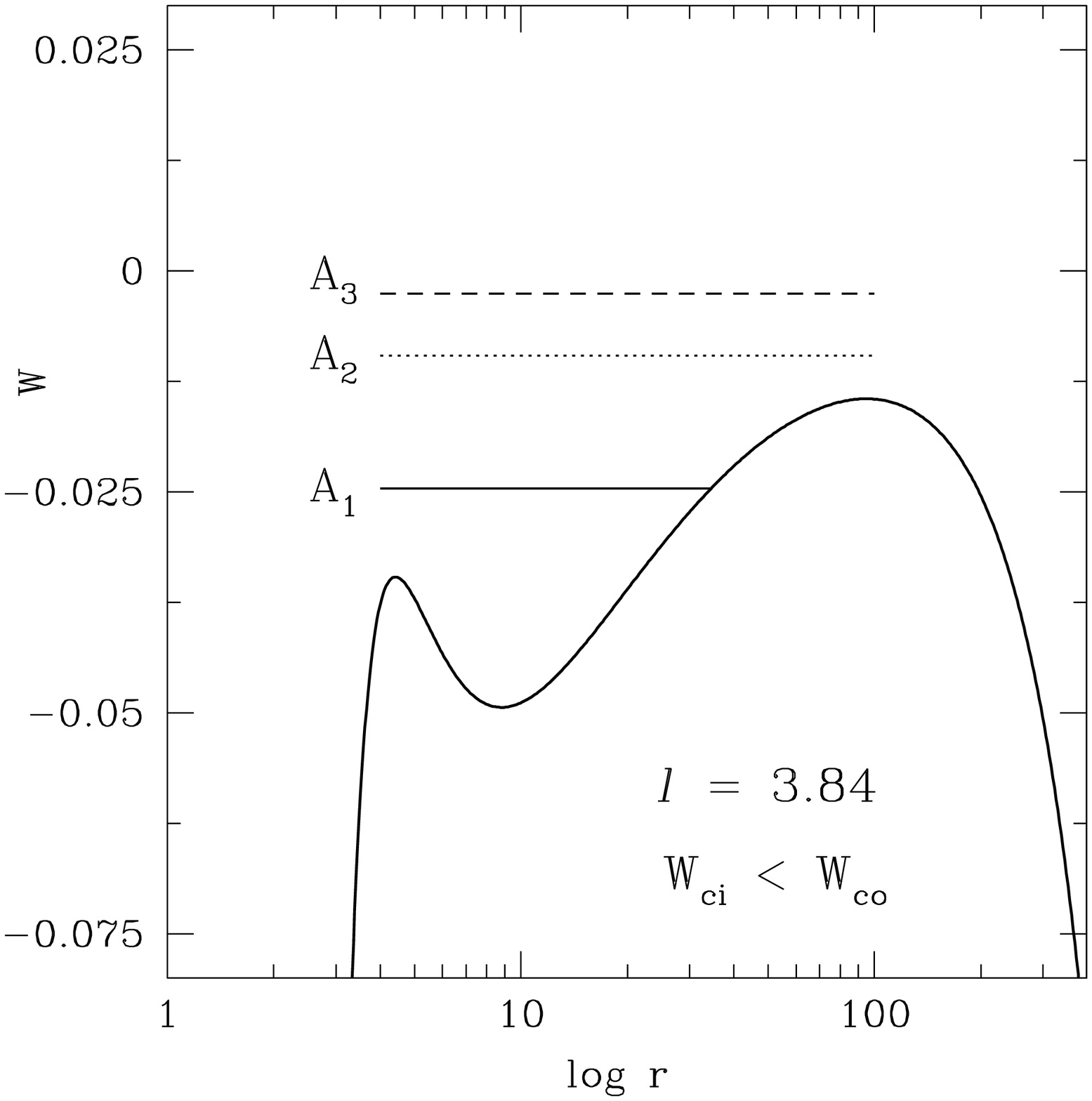}
\hspace{-1.45truecm}
\includegraphics[width=6.75cm,angle=0]{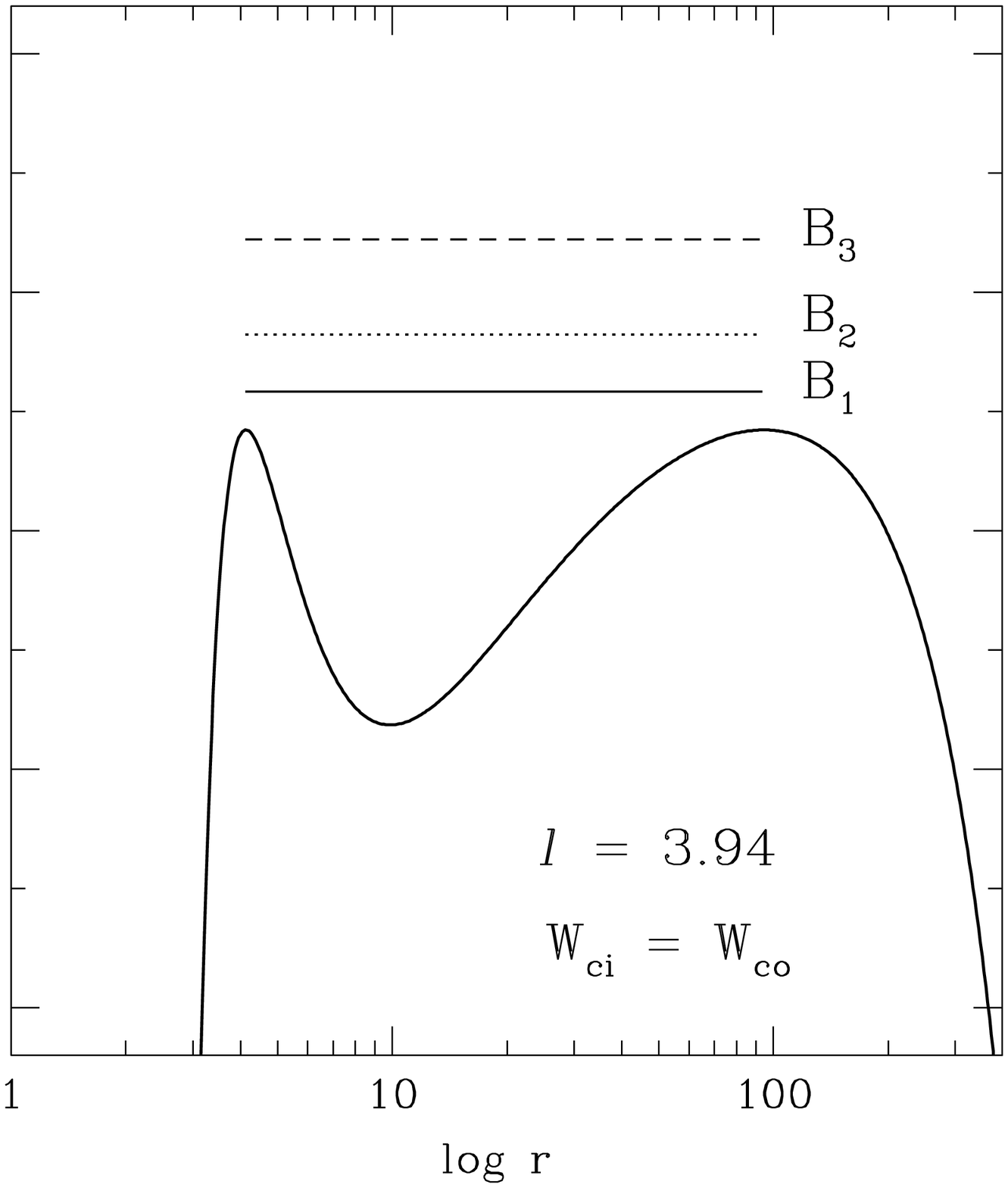}
\hspace{-1.1truecm}
\includegraphics[width=6.75cm,angle=0]{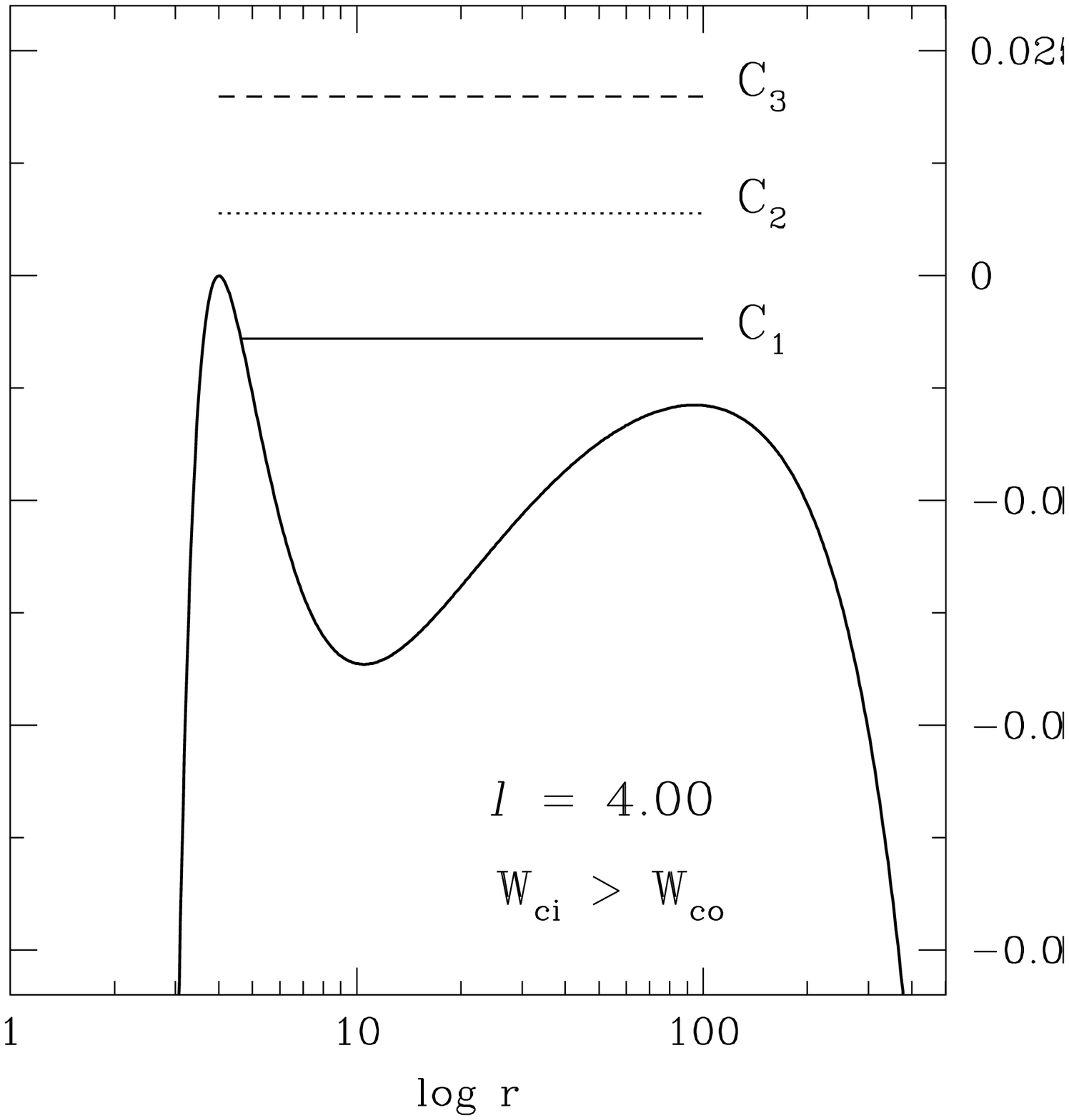}
\caption{Radial profiles of the potential at the equatorial plane for
models labelled $A$ (left), $B$ (centre) and $C$ (right),
respectively. The local maxima in each plot indicate the location of the
cusps. The horizontal lines fix the potential of the fluid element at the
edges of the torus for each of the models considered. Models $A$, $B$ and
$C$ differ among each other by the value of the angular momentum $\ell$
which is reported in the lower right corners of the different panels.  }
\label{fig1}
\end{center}
\end{figure*}
%

%%%%%%%%%%%%%%%%%%%%%%%%%%%%%%%
\subsection{Initial Data}
\label{IV}
%%%%%%%%%%%%%%%%%%%%%%%%%%%%%%%

	As shown by Font \& Daigne (2002a) and Zanotti {\rm et al.}
(2003), the runaway instability is a robust feature of constant angular
momentum relativistic tori in Schwarzschild and Kerr spacetimes, if these
are non self-gravitating. This result does not depend on the way the
instability is triggered, i.e. by either artificially expanding the torus
over the potential barrier at the inner cusp (Font \& Daigne, 2002a), or
by introducing perturbations in an otherwise stable torus (Zanotti {\it
et al.}  2003).  As discussed above, however, the presence of an outer
cusp in a Schwarzschild-de Sitter spacetime is likely to affect the
robustness of this conclusion. In order to investigate to what extent an
outflow of mass can interfere with the development of the runaway
instability we have studied the behaviour of three different classes of
models, which we refer to as $A$, $B$ and $C$. These models are
distinguished on whether the effective potential at the inner cusp,
$W_{\rm{ci}}$, is less than, equal to, or larger than the effective
potential $W_{\rm{co}}$ at the outer cusp, respectively. Furthermore, for
each of these classes of models we have considered three different
initial configurations, as shown in Fig.~\ref{fig1}, with the potential
at the inner edge of the torus, $W_{\rm{in}}$, being different from the
potential barrier, $W_{\rm{ci}}$, by the adjustable amount $\Delta
W_{\rm{i}}$.

	Note that for all of the models of the class $A$, the hydrostatic
equilibrium is always violated at the inner cusp, i.e.  $W_{\rm {in}} >
W_{\rm{ci}}$, and a mass outflow will necessarily take place at the inner
edge of the disc once the initial data is evolved. Furthermore, a mass
loss will take place also at the outer cusp for models $A_2$ and $A_3$,
which have $W_{\rm {in}} > W_{\rm{co}} > W_{\rm{ci}}$. For all of the
models of the class $B$, on the other hand, the hydrostatic equilibrium
is violated at both cusps and by the same amount, i.e. $W_{\rm {in}} >
W_{\rm{ci}} = W_{\rm{co}}$ and, again, the mass outflows are regulated by
the potential jump $\Delta W_{\rm i}$. Finally, for all of the models of
the class $C$, the hydrostatic equilibrium is always violated at the
outer cusp, i.e. $W_{\rm {in}} > W_{\rm{co}}$, and also at the inner cusp
for models $C_2$ and $C_3$, for which $W_{\rm {in}} > W_{\rm{ci}} >
W_{\rm{co}}$. All of these different initial conditions are illustrated
in Fig.~\ref{fig1}, whose different panels show the effective potential
curves and the values of $W_{\rm {in}}$ for the three classes of models.

	Summarized in Table~\ref{tab1} are the potential jumps $\Delta
W_{\rm{i}}$ and $\Delta W_{\rm{o}}$, as well as the the most relevant
parameters of the various initial models considered here. Note that we
have used a polytropic index $\gamma=4/3$ and adjusted the polytropic
constant so as to have a small torus-to-hole mass ratio $M_{\rm
t}/M=0.2$, thus minimizing the error introduced by neglecting the
self-gravity of the torus.

	A final comment will be made on the value adopted for the
dimensionless cosmological constant $y$. As mentioned in the
Introduction, the primary goal of our investigation is to provide a first
qualitative analysis of the dynamics of thick discs in spacetimes with a
positive cosmological constant. Because of this, and in order to avoid
shortcomings with observations or computational resources, we have
performed numerical simulations with $y=10^{-6}$ for a black hole of mass
$M=10M_{\odot}$. These parameters, although unrealistic, allow us to
construct sufficiently compact discs and perform numerical simulations
with a satisfactory spatial resolution. We expect that all of the
qualitative conclusions we draw from our simulations will continue to
hold when considering more realistic values for both the cosmological
constant and the black hole masses.

%=======================================================
\section{Results}
\label{V}
%=======================================================

	As mentioned above, the development of the runaway instability
appears to be a robust feature of the dynamics of non self-gravitating
tori orbiting around Schwarzschild or Kerr black holes with constant
distributions of specific angular momentum. This conclusion has been
reached after numerous simulations have been performed for a large range
of torus-to-hole mass ratios $M_{\rm t}/M$ and under a number of
different initial conditions (Font \& Daigne, 2002a; Zanotti {\rm et al.}
2003). In all of these simulations, the onset and full development of the
instability was observed when the spacetime geometry was suitably
modified to account for the black hole's mass-increase due to
accretion. As a result of the instability, the torus is very rapidly
accreted onto the black hole and this is most clearly signalled by the
exponential growth in time of the rest-mass accretion rate at the
innermost radial grid point. The growth-time for the instability is
inversely proportional to the ratio $M_{\rm t}/M$ and is comparable with
the dynamical (i.e. orbital) timescale when $M_{\rm t}/M \sim 1$.

\begin{figure*}
\begin{center}
\includegraphics[width=8.7cm,angle=0]{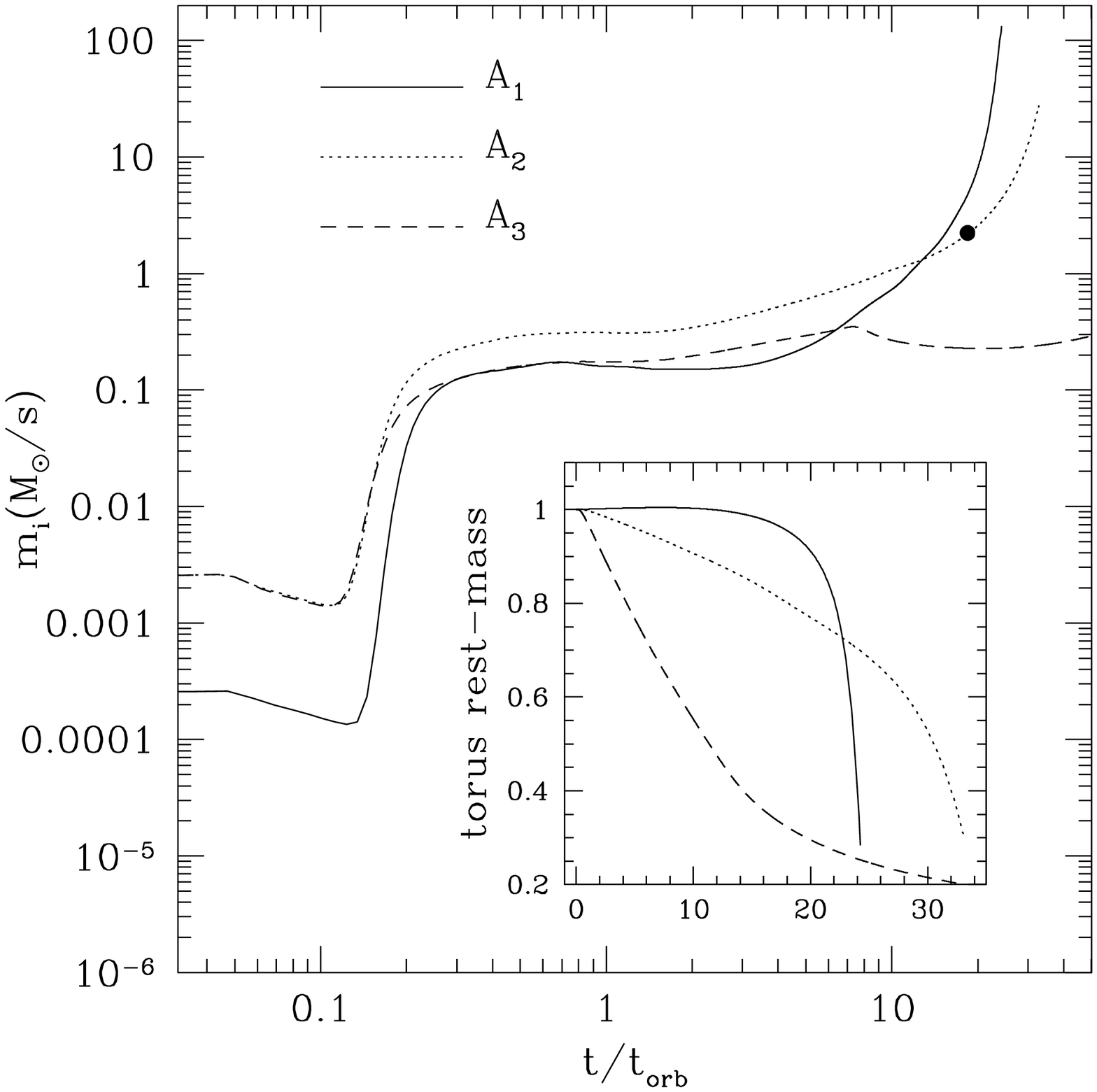}
\hspace{0.05truecm}
\includegraphics[width=8.7cm,angle=0]{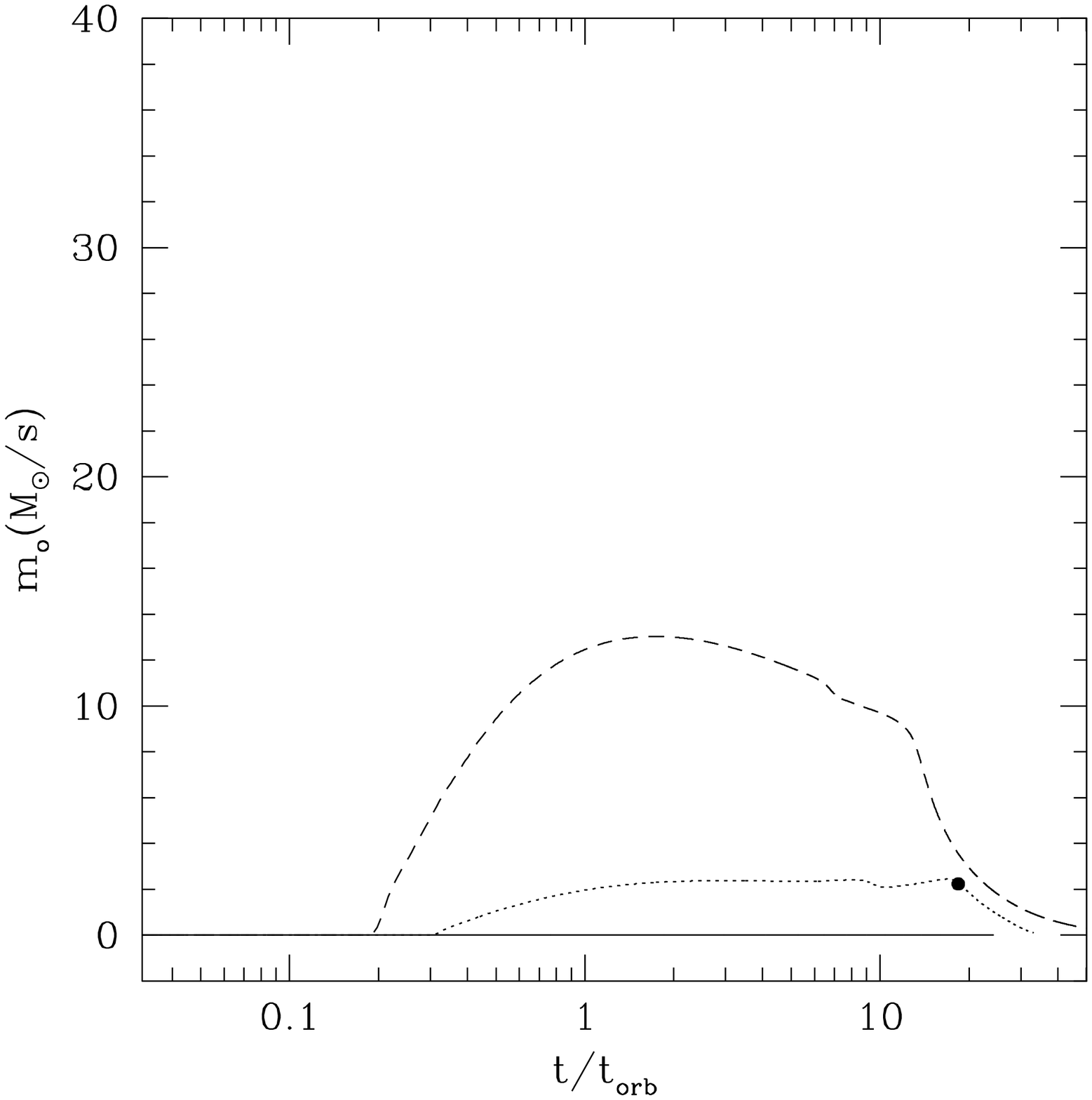}
\caption{Time evolution of the inner (left panel) and of the outer (right
panel) mass outflows for the models of class $A$. The data is shown in
units of solar masses per second, while the time is expressed in units
of the orbital period. Note that only models $A_1$ (solid line) and $A_2$
(dotted line) are runaway unstable. The solid circles in the two panels
indicate the time at which $\dot{m}_{\rm i} > \dot{m}_{\rm o}$ for model
$A_2$.}
\label{fig2}
\end{center}
\end{figure*}
\begin{figure*}
\begin{center}
\includegraphics[width=8.7cm,angle=0]{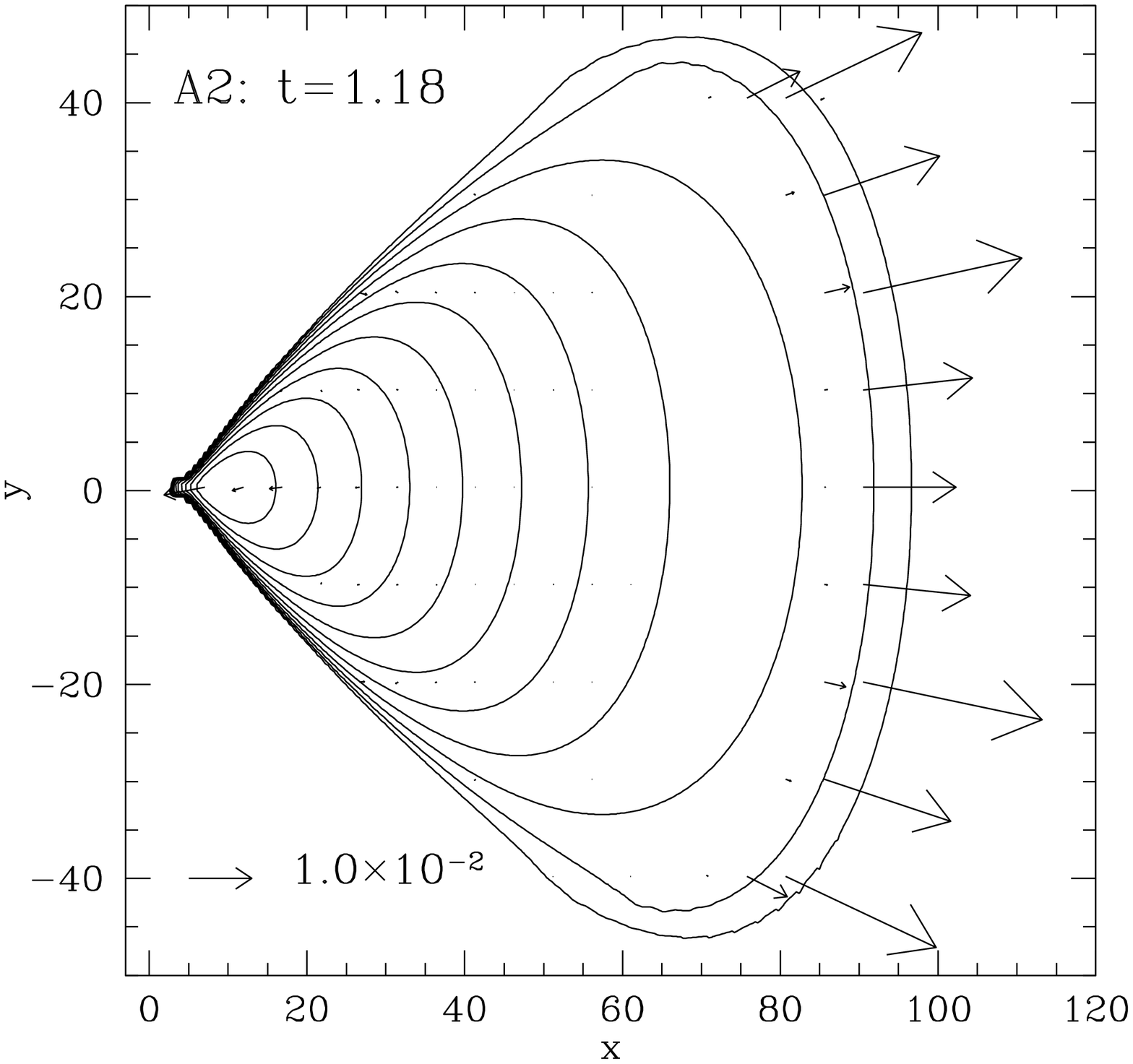}
\hspace{0.05truecm}
\includegraphics[width=8.7cm,angle=0]{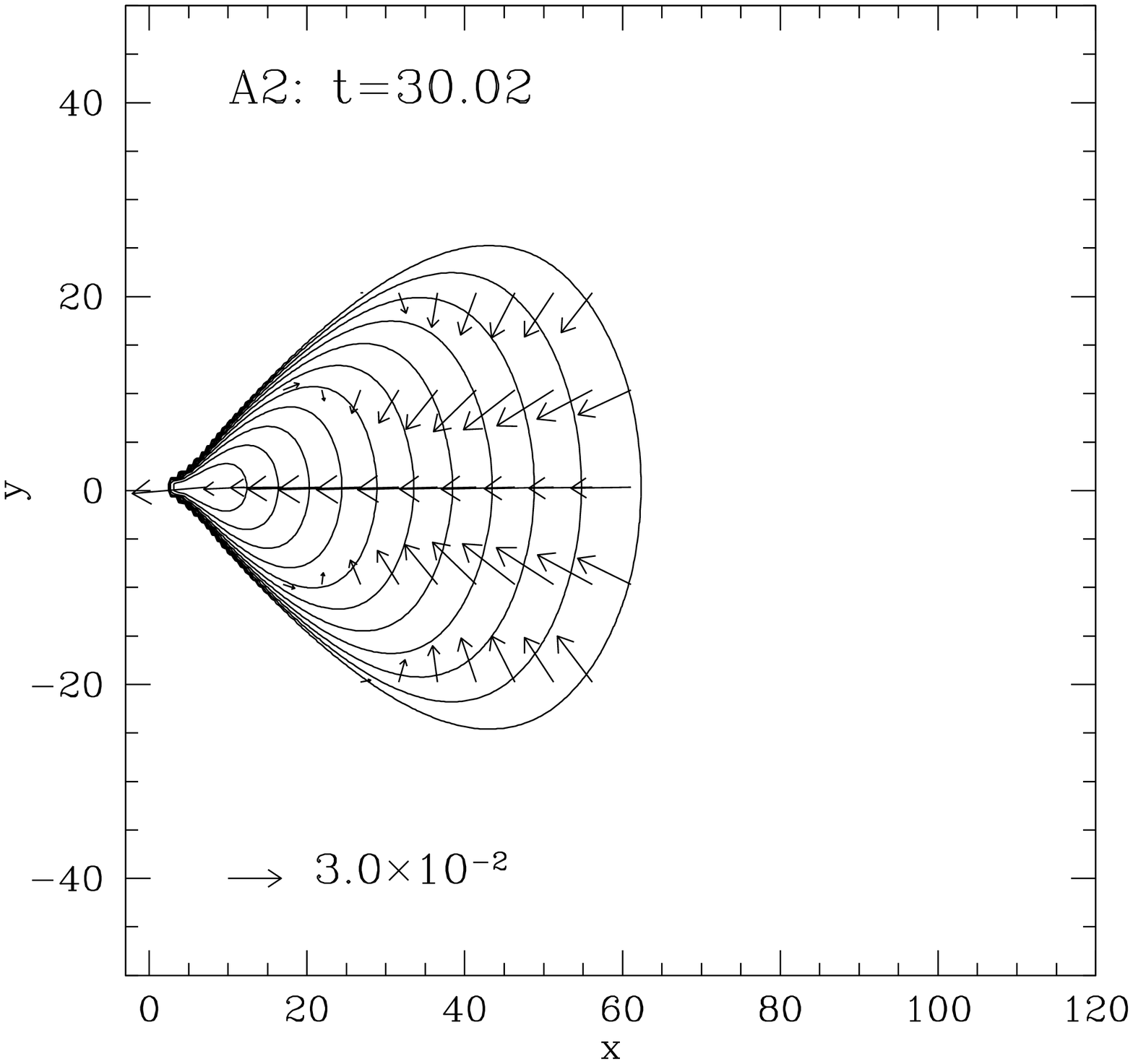}
\caption{Velocity field and equally spaced isocontours of the logarithm
of the rest-mass density for model $A_2$ at an early time (left panel)
and at a later time (right panel); the times reported are in units of the
orbital period. Initially the outer mass flux dominates the dynamics of
the torus. However, the gravitational attraction of the black hole
eventually overcomes the effect of the cosmological constant and the
runaway instability takes place. This leads to the large inward-directed
fluxes and to the disappearance of the torus inside the black hole in a
few orbital periods.}
\label{fig5}
\end{center}
\end{figure*}

	In the case of a Schwarzschild-de Sitter black hole, however, the
rest-mass of the torus can change not only because of losses through the
inner cusp leading to accretion onto the black hole, but also because of
outflows from the outer cusp and away from the black hole. While both
allowed, the impact that these two mass outflows could have on the
dynamics of the torus is very different. The first one, in fact, induces
changes in the black hole mass and could therefore lead to the runaway
instability. The second one, on the other hand, does not produce changes
of the background spacetime and cannot therefore produce an
instability. Nevertheless, it can affect the hydrodynamical evolution in
a number of different ways: firstly, by reducing the amount of rest-mass
in the torus available for accretion and, secondly, by producing
significant alterations of the velocity field, especially in the outer
regions of the torus.

\begin{figure*}
\begin{center}
\includegraphics[width=8.7cm,angle=0]{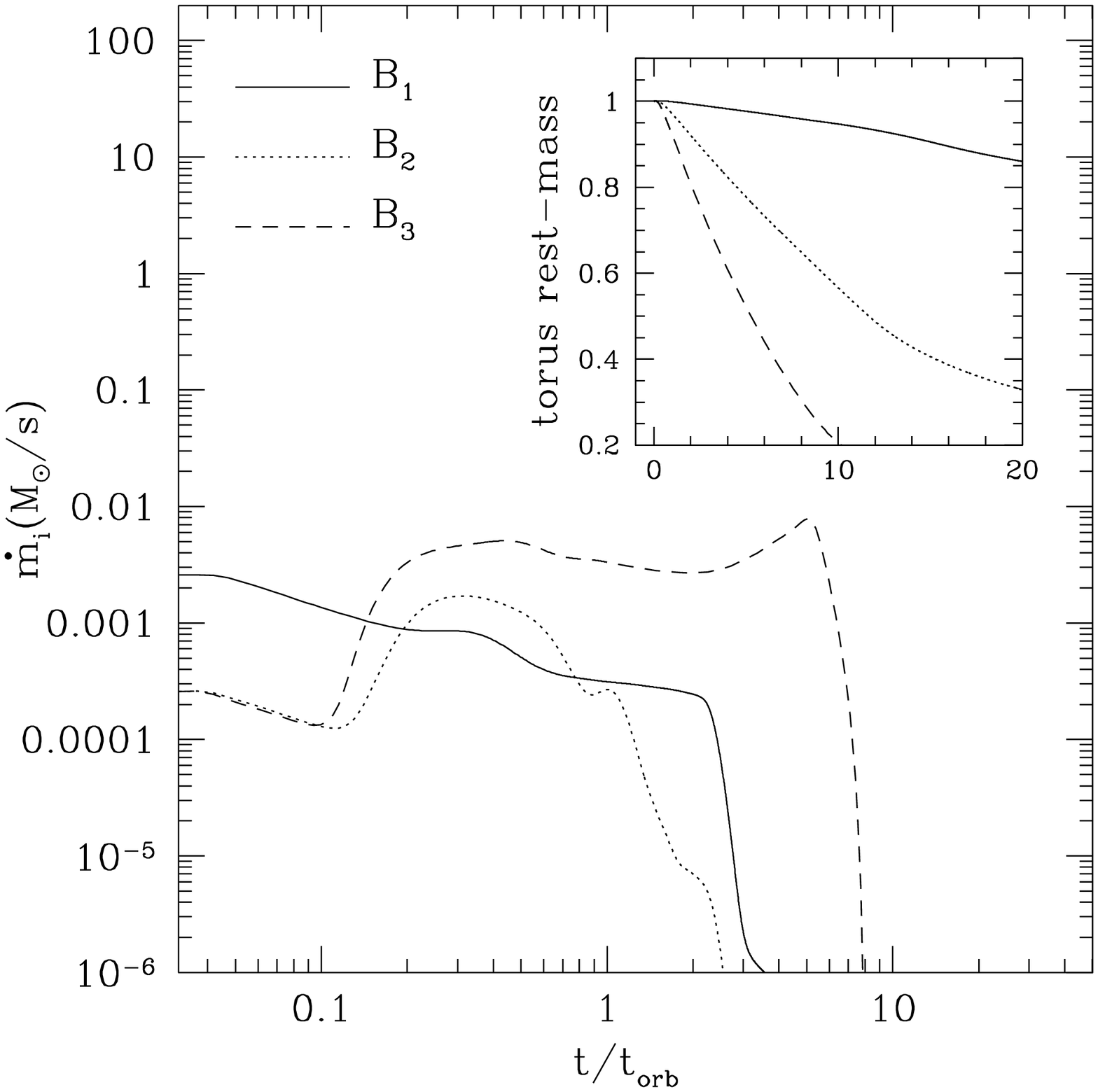}
\hspace{0.05truecm}
\includegraphics[width=8.7cm,angle=0]{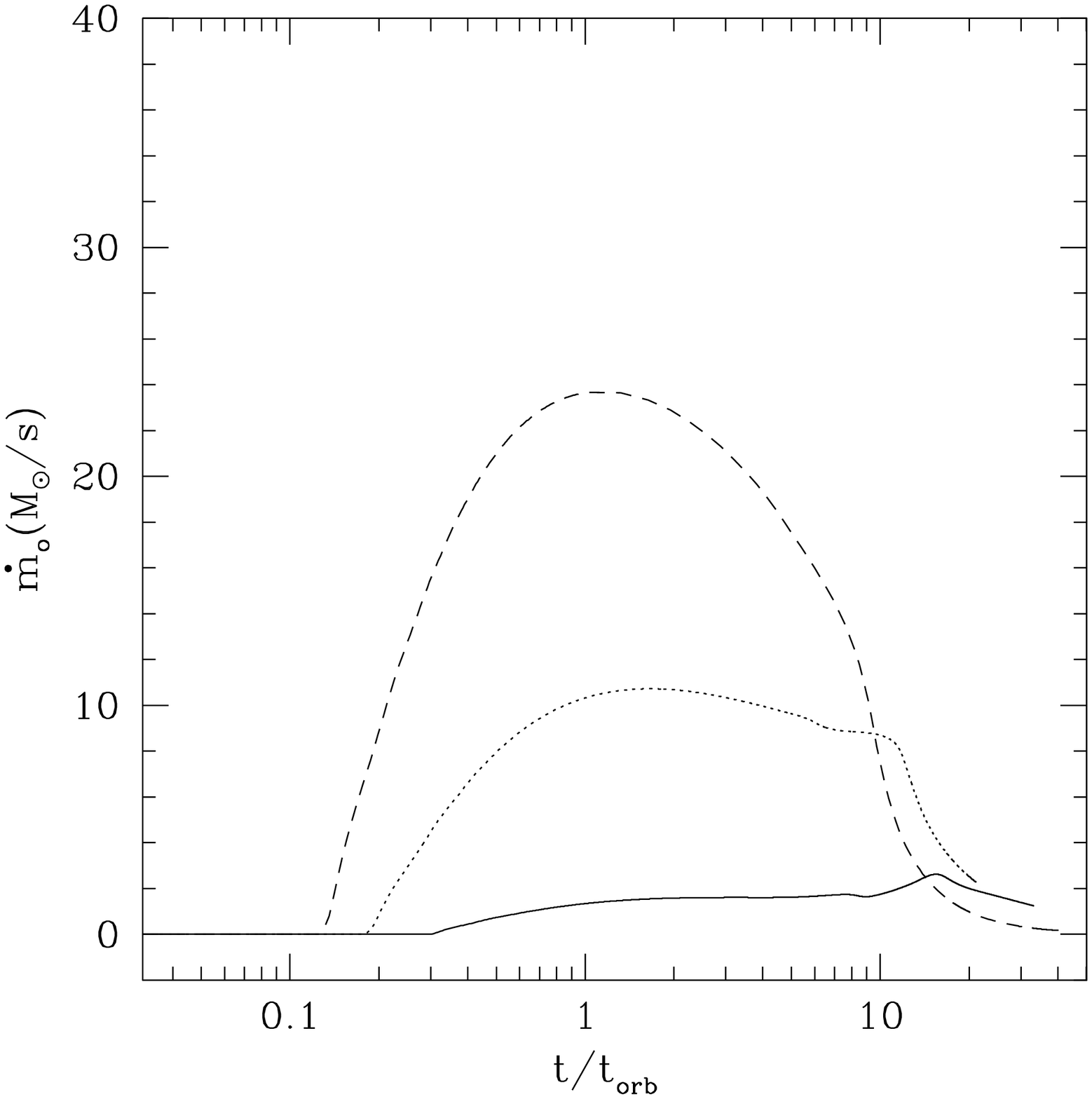}
\caption{Same as Fig.~\ref{fig2} but for the models of class $B$.}
\label{fig3}
\end{center}
\end{figure*}

\begin{figure*}
\begin{center}
\includegraphics[width=8.7cm,angle=0]{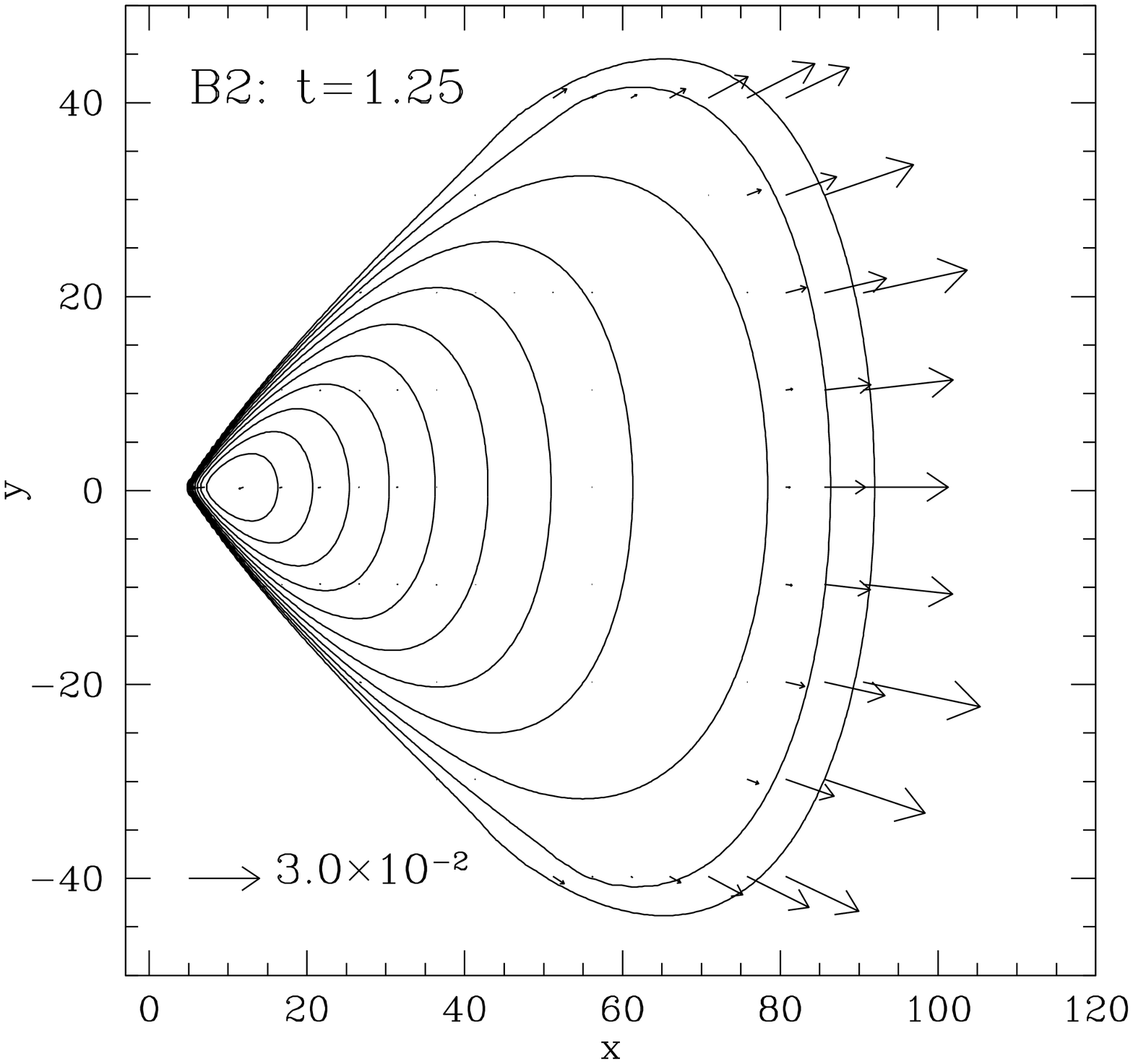}
\hspace{0.05truecm}
\includegraphics[width=8.7cm,angle=0]{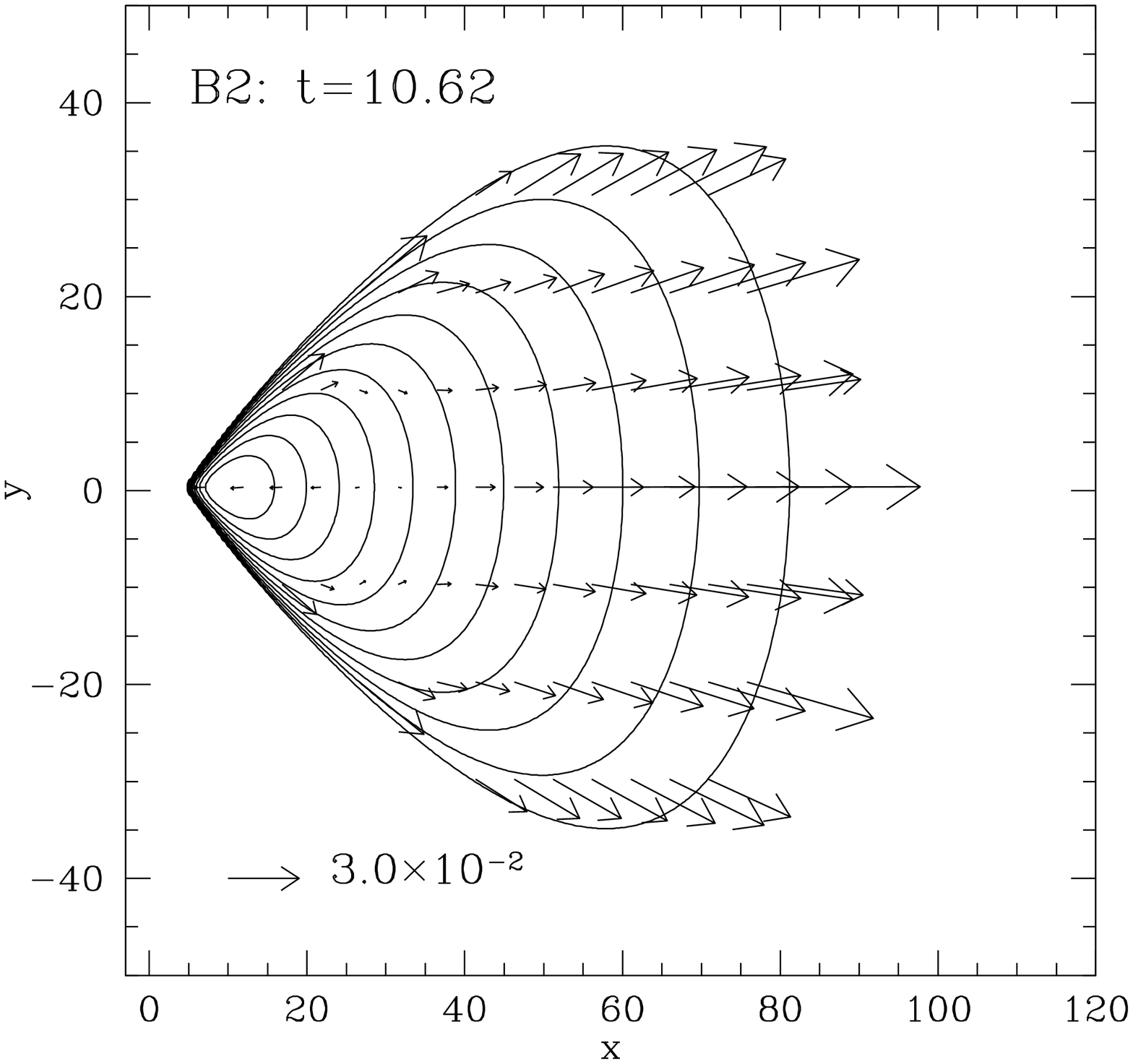}
\caption{Same as Fig.~\ref{fig5} but for model $B_2$. The intense mass
ouflow across the outer edge of the disc removes a large fraction of its
mass, and suppresses the runaway instability. The final disc reaches a
quasi-steady state.}
\label{fig6}
\end{center}
\end{figure*}

\begin{figure*}
\begin{center}
\includegraphics[width=8.7cm,angle=0]{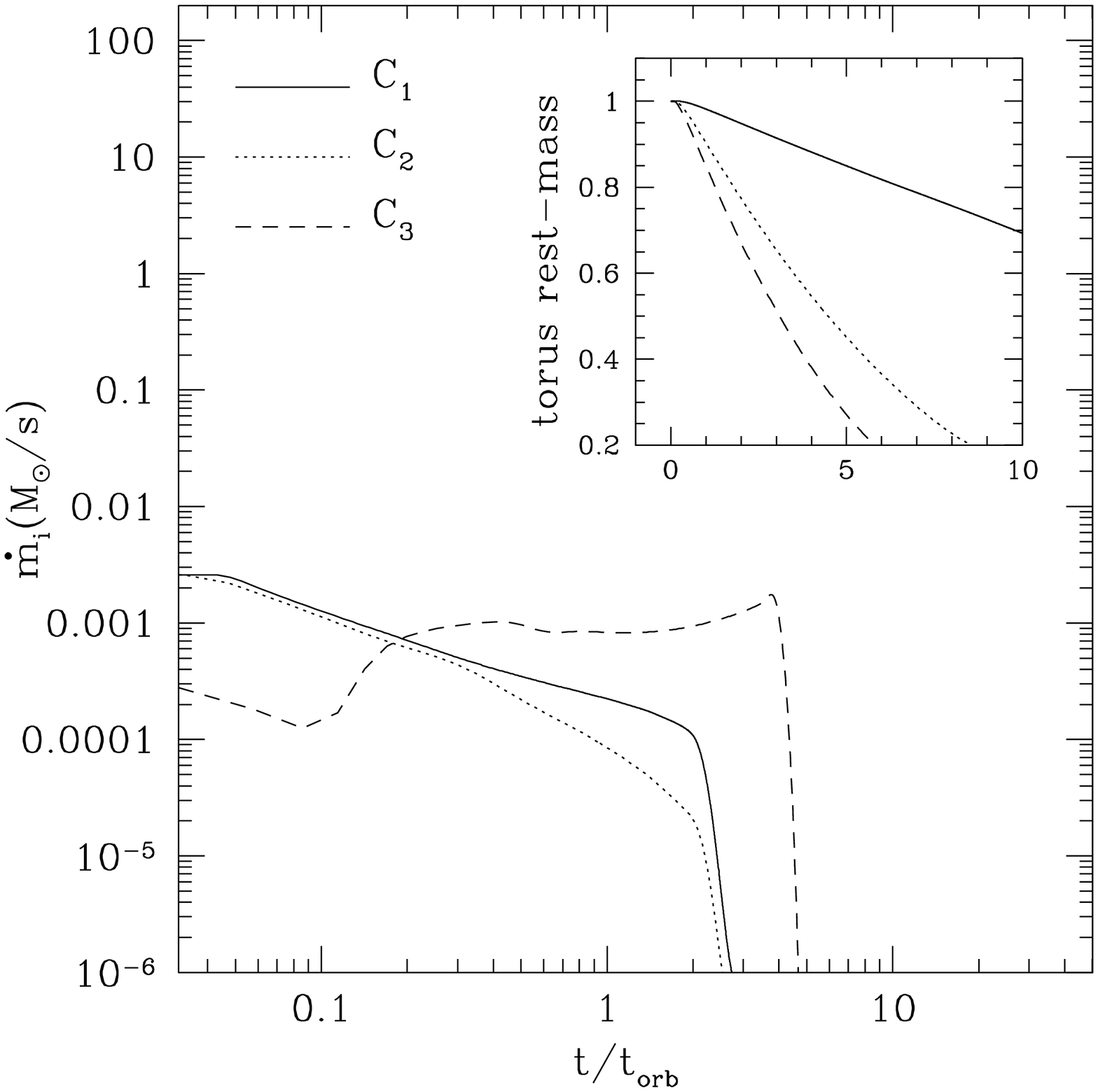}
\hspace{0.05truecm}
\includegraphics[width=8.7cm,angle=0]{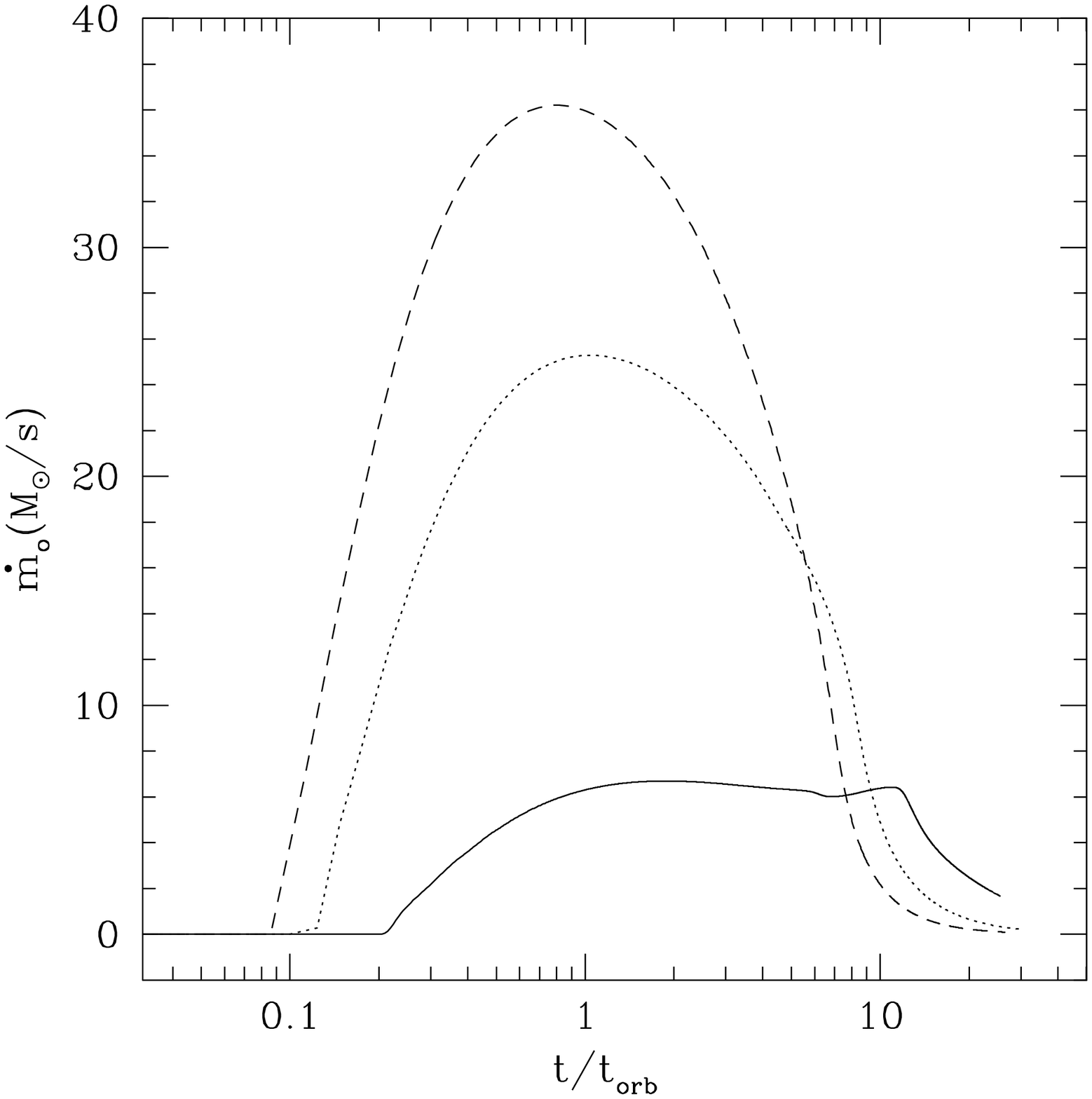}
\caption{Same as Fig.~\ref{fig2} but for the models of class $C$.}
\label{fig4}
\end{center}
\end{figure*}

	We have followed the hydrodynamical evolution of the models
described in Table~\ref{tab1} over a number of orbital periods
sufficiently large to reveal the impact of a positive cosmological
constant on the occurrence of the runaway instability. In
Figs.~\ref{fig2}, \ref{fig3} and \ref{fig4} we show the two mass outflows
$\dot{m}_{\rm i}$ and $\dot{m}_{\rm o}$ as a function of the orbital
period $t_{\rm orb}=2\pi/ \Omega_{\rm max}$ at the centre of the torus,
and for the three classes of models listed in Table~\ref{tab1}. The three
small insets shown in the panels for $\dot{m}_{\rm i}$ offer a view of
the evolution of the rest-mass of the torus after this has been
normalized to its initial value.  The description of the dynamics of the
tori is also completed with Figs.~\ref{fig5} and~\ref{fig6}, which show
equally spaced isocontours of the logarithm of the rest-mass density and,
superimposed, the velocity field for models $A_2$ and $B_2$,
respectively, at two different times during the evolution.

	As it is apparent from a rapid look at these figures, the runaway
instability is no longer the only possible evolution of the system, whose
dynamics is instead the result of the interplay between the inner and the
outer mass outflows. The occurrence of the runaway instability is clearly
visible in the left panel of Fig.~\ref{fig2} for model $A_1$ (solid
line). Model $A_1$, in fact, has initial conditions that resemble those
encountered for a Schwarzschild black hole, with the outer radius of the
torus located far from the outer cusp (cf. Figs.~\ref{fig0}--\ref{fig1}).
As a result, the right panel of Fig.~\ref{fig2}, shows that the outer
mass outflow is in this case very small (indeed slightly negative as a
result of accretion onto the torus of the infalling atmosphere), while
the mass accretion rate onto the black hole (left panel) grows
exponentially and undisturbed until the full development of the runaway
instability at $t\sim 24.2\; t_{\rm orb}$. The dynamical evolution is
different for models $A_2$ (dotted line) and $A_3$ (dashed line), where
the competition between the two outflows at the inner and outer edges of
the disc is closer to a balance and the initial outer mass-loss is
non-negligible. For model $A_2$, in particular, this is clearly visible
in the left panel of Fig.~\ref{fig5}, which shows that at early times
($t\sim 1.2\; t_{\rm orb}$) the largest fluid velocities ($v\sim 4\times
10^{-3}$) are reached in the outer regions of the torus and are outwardly
directed. However, the outer mass-loss eventually becomes insufficient to
prevent the development of the runaway instability, which takes place
after $t\sim 33\; t_{\rm orb}$. The corresponding velocity field (with
all vectors pointing towards the black hole) and the isocontours of the
logarithm of the rest-mass density displayed in the right panel of
Fig.~\ref{fig5} (at time $t\sim 30.0\; t_{\rm orb}$) show the important
reduction in size undergone by the torus, which is about to disappear
entirely inside the black hole after a few more orbital periods. This
type of evolution does not take place for model $A_3$, whose dynamics is
completely dominated by the mass outflow through the outer cusp and for
which the runaway instability does not develop (cf. Fig.~\ref{fig2}).


	To better interpret the dynamics behind these simulations it is
useful to compare the amount of rest-mass in the torus after the
first 10 orbital periods of the evolution for the three models of class
$A$ (see inset in the left panel of Fig.~\ref{fig2}). The residual
rest-mass of the torus is $99\%$, $90\%$, and $55\%$ of the initial one
for the models $A_1$, $A_2$, and $A_3$, respectively. This may appear
somewhat surprising given the fact that the mass outflow from the inner 
cusp is smaller in the case of model $A_3$ than it is for model $A_2$, 
despite $A_3$ having a larger initial potential jump $\Delta W_{\rm i}$. 
The explanation for this comes from looking at the right panel of
Fig.~\ref{fig2} which shows that the mass outflow from the outer cusp is
however larger for model $A_3$ than it is for model $A_2$. As a result,
the torus is emptied more efficiently, and this happens mostly through
the outer cusp.

	It is also very instructive to compare the mass outflows at the
inner and outer edges of the disc for the three models of class
$A$. Doing so leads to the important result that $\dot{m}_{\rm i} >
\dot{m}_{\rm o}$ at all times for model $A_1$ and that $\dot{m}_{\rm i} <
\dot{m}_{\rm o}$ at all times for model $A_3$. In other words, the
unstable and the stable models seem to differ from each other on whether
the mass outflow from the inner edge is larger or smaller than the
corresponding mass-loss from the outer edge. In the case of model $A_2$,
on the other hand, the two mass outflows are closer to a balance and
$\dot{m}_{\rm i} > \dot{m}_{\rm o}$ only for $t \gtrsim 18 \; t_{\rm
orb}$, after which the runaway instability clearly develops; the time
when this happens is shown with the filled circles in the two panels of
Fig.~\ref{fig2}.  It appears, therefore, that an increasing potential gap
at the outer edge of the torus {\it favours} the outflow of mass from the
{\it outer} cusp but it also {\it disfavours} (and to a larger extent)
the mass outflow from the {\it inner} cusp, with the corresponding
accretion onto the black hole being severely suppressed. As a result, the
feed-back of the black hole spacetime onto the overall dynamics of the
system is considerably reduced and the runaway instability
suppressed. Stated differently, the development of the runaway
instability appears to be related to the efficiency of the mass-loss
through the edges of the disc and, in particular, the instability is
effectively ``extinguished'' whenever the mass outflow from the {\it
outer edge} of the disc and away from the black hole is {\it larger} than
the mass outflow from the {\it inner edge} of the disc and onto the black
hole.  The condition $\dot{m}_{\rm i} < \dot{m}_{\rm o}$ can thus be used
as a simple sufficient condition for the suppression of the runaway
instability in a thick disc orbiting around a Schwarzschild-de Sitter
black hole.

	The role played by a positive cosmological constant on the
dynamics of the discs and described so far for the models of class $A$ is
present also for the models of class $B$ and $C$, although with some
slight differences. As discussed in Sect.~\ref{IV} and illustrated in
Fig.~\ref{fig1}, the models of class $B$ are built with the outer and
inner edges having the same effective potential. One would therefore
expect that this would yield to very similar mass outflows at the two
boundaries of the disc. However, the two panels of Fig.~\ref{fig3} show
that the mass fluxes through the outer edges of the discs to infinity are
always larger than the ones towards the black hole (i.e. $\dot{m}_{\rm i}
< \dot{m}_{\rm o}$). As a result, the models of class $B$ are all stable
to the runaway instability. This is particularly apparent in models $B_2$
and $B_3$, for which the outer mass outflows are at least a couple of
orders of magnitude larger than the corresponding mass outflows onto the
black hole, and which become negligibly small (i.e. ${\dot m}_{\rm i} <
10^{-8} M_{\odot}/{\rm s}$) well before 10 orbital periods.  As a result,
a large amount of the matter in those discs is not accreted onto the
black hole, but escapes to infinity. This is illustrated in the small
inset of the left panel of Fig.~\ref{fig3} which shows that after about
20 orbital periods more than $60\%$ of the torus rest-mass is lost for
model $B_2$ and more than $90\%$ for model $B_3$.

	Once the outflows die-off in the tori of class $B$, the remaining
matter reaches a quasi-stable equilibrium, accreting onto the black hole
on a timescale which is essentially controlled by the rate of mass-loss
through the inner cusp. The importance of the mass outflow at the outer
edges of the discs of class $B$ is also apparent from Fig.~\ref{fig6},
which shows the velocity field and isocontours of the logarithm of the
rest-mass density of model $B_2$ at an early and a later stage of the
evolution. Note how the left panels of Fig.~\ref{fig5} and \ref{fig6}
have velocity fields that differ mostly in modulus but are equally
oriented, while the right panels are substantially different with
velocity fields that have opposite orientations leading to the
disappearence of the torus into the black hole and to infinity,
respectively. We also note that while the difference between the inner
and outer mass outflows remains large also in the case of model $B_1$,
the dynamics is in this case much closer to an equilibrium, with the
torus being still progressively emptied to infinity, but on a much larger
timescale. No runaway instability was observed for this model over the
time for which the calculations were carried out ($t \sim 33\; t_{\rm
orb}$).

	Finally, for models $C$ (see Fig.~\ref{fig4}), the dynamics of
the discs is particularly simple and the final result is rather clear to
interpret. In this case, in fact, all the discs are built with an
effective potential which is larger at the outer edge
(cf. Fig.~\ref{fig1}) and represent, therefore, initial conditions that
are conceptually the opposite of those in models $A$. Because of the high
potential barrier at the inner edge of the disc, the inner mass outflow
is always rather minute and several orders of magnitude smaller than the
corresponding mass outflow from the outer edge. As a result, the mass in
the torus is lost very rapidly to infinity and very little is accreted
onto the black hole. In particular, in the most dramatic case of model
$C_3$, the residual rest-mass in the torus is less than $20\%$ after only
6 orbital periods.

%=======================================================
\section{Conclusions}
\label{VI}
%=======================================================

	We have investigated the effect of a positive cosmological
constant on the dynamics of non self-gravitating thick accretion discs
orbiting Schwarzschild-de Sitter black holes with constant distributions
of specific angular momentum. The motivation behind this investigation
has been that of assessing the role played by an effective repulsive
force in the onset and development of the runaway instability, which
represents a robust feature in the dynamics of constant angular momentum
tori. In addition to the inner cusp near the black hole horizon, through
which matter can accrete onto the black hole when small deviations from
the hydrostatic equilibrium are present, thick discs in a Schwarzschild
de-Sitter spacetime also possess an outer cusp through which matter can
leave the torus without accreting onto the black hole. As a result of
this mass-loss to infinity, the changes in the background metric (which
are responsible for the development of the runaway instability) may be
altered considerably and the instability thus suppressed.

	As a simple way to evaluate this effect we have considered a
sequence of Schwarzschild-de Sitter spacetimes differing only in their
total mass and have performed time-dependent general relativistic
hydrodynamical simulations in these background metrics of thick discs
which are initially slightly out of hydrostatic equilibrium. In doing
this we have adopted an unrealistically high value for the cosmological
constant which however yields sufficiently small discs (extending up to
about a few hundred gravitational radii) to be accurately resolved with
fine enough axisymmetric numerical grids.

	We have performed a number of simulations involving initial
configurations of constant specific angular momentum discs differing both
for the relative amplitude of the peaks in the effective potential and
for the potential jump at the inner and outer cusps. The results obtained
indicate that the runaway instability is no longer the only possible
evolution of these systems but that their dynamics is rather the
end-result of the interplay between the inner and the outer mass
outflows. On the one hand, in fact, we have evolved initial models for
which the cosmological constant has a weak influence; these models have
negligible mass outflows to infinity while maintaining large mass
outflows onto the black hole, which then lead to the development of the
runaway instability. On the other hand, we have evolved initial models
which are significantly influenced by the cosmological constant; these
models develop mass outflows through the outer cusp which are much larger
than those appearing at the inner cusp and, hence, do not develop the
runaway instability. Placed somewhere between these two classes of
initial configurations there exist initial models for which the mass
outflows from the inner and outer cusps are more closely balanced. In
these cases the runaway instability may or may not develop and we have
noticed that a simple comparison between the mass outflows can be used to
deduce the fate of the accreting disc. More specifically, we have found
that the condition $\dot{m}_{\rm i} < \dot{m}_{\rm o}$ provides a simple
sufficient condition for the suppression of the runaway instability in a
thick disc orbiting around a Schwarzschild-de Sitter black hole.

	In spite of the idealized setup used, the simulations performed
here provide a first qualitative description of the complex nonlinear
dynamics of thick discs in Schwarzschild-de Sitter spacetimes and we
expect that most of the results obtained will continue to hold also when
more realistic values for the cosmological constant are used. Aa a final
comment we note that besides providing a qualitative description of the
role that a cosmological constant could play on the dynamics of
relativistic tori, these calculations also offer a way of assessing, at
least qualitatively, the inertial role that the self-gravity of the torus
plays in the development of the runaway instability. This will be very
useful when studying the dynamics of relativistic tori with numerical
codes solving also the full Einstein equations.

%=======================================================
\begin{acknowledgements}

It a pleasure to thank Z. Stuchl\'{\i}k for useful comments. Financial
support for this research has been provided by the Italian MIUR, by the
Spanish Ministerio de Ciencia y Tecnolog\'{\i}a (grant AYA
2001-3490-C02-01) and and by the EU Network Programme (Research Training
Network Contract HPRN-CT-2000-00137). LR also acknowledges the kind
hospitality at the KITP in Santa Barbara, where part of this research was
carried out (NSF grant PHY99-07949). The computations were performed on
the Beowulf Cluster for numerical relativity {\it ``Albert100''}, at the
University of Parma.

\end{acknowledgements}
%=======================================================

\end{document}